%% file: full_article.tex
\documentclass{aastex62}

\usepackage{hyperref}
\usepackage{listings}
\usepackage{amsmath}
\usepackage[top=1.5in,bottom=0.5in,left=1.25in,right=0.65in]{geometry}

\newcommand{\params}{\boldsymbol{\theta}}
\newcommand{\cparams}{\boldsymbol{\alpha}}

\shortauthors{Zucker et al.}

\begin{document}

\title{A Large Catalog of Accurate Distances to Local Molecular Clouds: The Gaia DR2 Edition}

\correspondingauthor{Catherine Zucker}
\email{catherine.zucker@cfa.harvard.edu}

\author{Catherine Zucker}
\altaffiliation{Equal contribution}
\affil{Harvard Astronomy, Harvard-Smithsonian Center for Astrophysics, 60 Garden St., Cambridge, MA 02138, USA}

\author{Joshua S. Speagle}
\altaffiliation{Equal contribution}
\affil{Harvard Astronomy, Harvard-Smithsonian Center for Astrophysics, 60 Garden St., Cambridge, MA 02138, USA}

\author{Edward F. Schlafly}
\affil{Lawrence Berkeley National Laboratory, One Cyclotron Road, Berkeley, CA 94720, USA}

\author{Gregory M. Green}
\affil{Kavli Institute for Particle Astrophysics and Cosmology, Physics and Astrophysics Building, 452 Lomita Mall, Stanford, CA 94305, USA}

\author{Douglas P. Finkbeiner}
\affil{Harvard Astronomy, Harvard-Smithsonian Center for Astrophysics, 60 Garden St., Cambridge, MA 02138, USA}

\author{Alyssa A. Goodman}
\affil{Harvard Astronomy, Harvard-Smithsonian Center for Astrophysics, 60 Garden St., Cambridge, MA 02138, USA}
\affil{Radcliffe Institute for Advanced Study, Harvard University, 10 Garden St, Cambridge, MA 02138}

\author{Jo{\~a}o Alves}
\affil{Radcliffe Institute for Advanced Study, Harvard University, 10 Garden St, Cambridge, MA 02138}
\affil{University of Vienna, Department of Astrophysics, T{\"u}rkenschanzstra{\ss}e 17, 1180 Vienna, Austria}



\begin{abstract}
We present a uniform catalog of accurate distances to local molecular clouds informed by the Gaia DR2 data release. Our methodology builds on that of \citet{Schlafly_2014}. First, we infer the distance and extinction to stars along sightlines towards the clouds using optical and near-infrared photometry. When available, we incorporate knowledge of the stellar distances obtained from Gaia DR2 parallax measurements. We model these per-star distance-extinction estimates as being caused by a dust screen with a 2-D morphology derived from Planck at an unknown distance, which we then fit for using a nested sampling algorithm. We provide updated distances to the \citet{Schlafly_2014} sightlines towards the \citet{Dame_2001} and \citet{Magnani_1985} clouds, finding good agreement with the earlier work. For a subset of 27 clouds, we construct interactive pixelated distance maps to further study detailed cloud structure, and find several clouds which display clear distance gradients and/or are comprised of multiple components. We use these maps to determine robust average distances to these clouds. The characteristic combined uncertainty on our distances is $\approx 5-6\%$, though this can be higher for clouds at farther distances, due to the limitations of our single-cloud model. 
\end{abstract}




\section{Introduction} \label{sec:intro}

Stars are the fundamental unit of astronomy, acting as both the hosts of planetary systems and the building blocks of galaxies. Understanding how stars form underpins our understanding of physics from planetary to galactic scales. Due to the close connection between molecular gas and star formation, molecular clouds have been targeted in an effort to understand the process by which dense molecular gas gravitationally collapses to form stars. Observations of molecular clouds in the solar neighborhood offer the best spatial resolution and facilitate both detailed studies of their physical properties and the testing of star formation theory. 

Accurate estimates of cloud mass, physical size, and star formation, however, depend critically on accurate distance measurements. Although there is a long history of using star colors and magnitudes to determine cloud distances, until recently distance estimates to local molecular clouds were obtained on a cloud-by-cloud basis using a variety of techniques with different (and often large) uncertainties. \citet{Schlafly_2014} took advantage of the Pan-STARRS1 survey \citep{Chambers_2016}, which observed the entire sky north of a declination $\delta=-30^\circ$, to produce a homogeneous catalog of accurate distances to local molecular clouds. In addition to most of the clouds in the \citet{Magnani_1985} catalog (hereafter MBM), \citet{Schlafly_2014} provided distances to well-studied molecular clouds like Orion, $\lambda$ Orionis, Taurus, Perseus, California, Ursa Major, the Polaris Flare, the Cepheus Flare, Lacerta, Pegasus, Hercules, Camelopardis, Ophiuchus, and Monoceros R2.

\citet{Schlafly_2014} determined the distances to molecular clouds in a two step process. First, they inferred the joint probability distribution function on distance and reddening for individual stars based on their optical photometry from Pan-STARRS1 following \citet{Green_2014}. Then, by modeling the cloud as a simple dust screen with a spatial template given by Planck \citep{Planck_2014}, they bracketed the dust screen between unreddened foreground stars and reddened background stars, thereby constraining at which distance a single ``jump'' in reddening occurs. 

In the last few years, there have been a variety of scientific developments that offer the potential for substantially more accurate distances to local molecular clouds. First, recent astrometric results from the Gaia DR2 release \citep{Lindegren_2018} offer an unprecedented opportunity to improve the accuracy and precision of this method, since stellar parallax measurements provide an independent (and often superior) constraint on stellar distances. In addition, new results from APOGEE allow data-driven modeling of possible variation of the optical-NIR extinction curve \citep{Schlafly_2016}. Finally, larger samples and improved modeling allow for more systematic exploration of the spatial structure of these clouds.


In \S \ref{sec:data} we describe the optical, near-infrared, and astrometric data on which our results are based, along with quality cuts we impose to obtain a reliable stellar sample. In \S \ref{sec:method} we describe our method for obtaining distances, including our updated framework for inferring the per-star distance extinction estimates and our line-of-sight dust model. In \S \ref{sec:results} we present an updated catalog of distances for the \citet{Schlafly_2014} sightlines, a new catalog of pixelated maps systematically covering the same regions, and a collection of average distances to each cloud. In \S \ref{sec:discussion} we compare our findings with other results from the literature and discuss the limitations of our model. We conclude in \S \ref{sec:conclusion}.

\section{Photometric and Astrometric Data} \label{sec:data}

We utilize data from four surveys: the Pan-STARRS1 Survey \citep[PS1;][]{Chambers_2016}, the National Optical Astronomy Observatory (NOAO) Source Catalog \citep[NSC;][]{nidever_18}, the Two Micron All Sky Survey \citep[2MASS;][]{Skrutskie_2006}, and the second data release of the Gaia survey \citep[Gaia DR2;][]{Brown_2018}. While the PS1 survey was previously utilized in \citet{Schlafly_2014}, the incorporation of NSC, 2MASS, and Gaia DR2 is novel to this work.

\subsection{Pan-STARRS1} \label{subsec:ps1}

The PS1 survey is a multi-epoch, deep broadband optical survey of the northern sky visible from Haleakala in Hawaii ($\delta > -30 ^\circ$). It observed in five photometric bands ($g_{\rm PS1}$, $r_{\rm PS1}$, $i_{\rm PS1}$, $z_{\rm PS1}$, $Y_{\rm PS1}$) spanning $400-1000$ nm, with a typical single epoch $5\sigma$ point-source exposure depth of 22.0, 21.8, 21.5, 20.9, and 19.7 magnitudes, respectively, in the \textit{AB system} \citep{okegunn_83}.

The photometry we utilize is based on catalog co-adds of single epoch photometry obtained as part of the PS1 DR1 $3\pi$ Steradian Survey. Compared to \citet{Schlafly_2014}, this work utilizes an additional 1.5 years of data from PS1, incorporates improved photometric and astrometric calibrations \citep{Magnier_2016}, and includes data from the north equatorial pole.

\subsection{NOAO Source Catalog} \label{subsec:nsc}

The NOAO Source Catalog is a catalog of sources derived from most of the public data taken on the NOAO telescopes, including the Dark Energy Camera (DECam). As this includes facilities in both the north and the south, the NOAO source catalog is the only dataset available that covers several local molecular clouds in the southern sky. Although the data are collected over many individual PI-led projects, the data were reduced and measured uniformly based on the NOAO Community Pipeline. The photometry is measured to roughly 1-2\% accuracy, with an astrometric accuracy of around 2 mas.

To ensure homogeneity, we only utilize NSC data taken from DECam in this analysis. These have a median 10$\sigma$ point-source exposure depth in $g_{\rm DECam}$, $r_{\rm DECam}$, $i_{\rm DECam}$, $z_{\rm DECam}$, and $Y_{\rm DECam}$ of 22.4, 22.9, 22.6, 22.4, 21.6, 20.4 magnitudes, respectively, in the \textit{AB system}. 

Although the data are reduced uniformly, because the number, duration, and general observing standards for exposures is heterogeneous over a given region of the sky, the overall quality of the data are worse than for PS1 (which had a uniform observing strategy across most of the main $3\pi$ survey). As a result, we only opt to use these data when no PS1 data are available, and will specifically highlight this in the remainder of the paper when relevant.

\subsection{2MASS} \label{subsec:2mass}

The 2MASS survey is a near-infrared (NIR) survey of the entire sky spanning $1-2 \micron$. It is shallower than PS1, achieving a typical $10\sigma$ point-source exposure depth in the $J_{\rm 2MASS}$, $H_{\rm 2MASS}$, and $K_{\rm 2MASS}$ bands of 15.8, 15.1, and 14.3 magnitudes, respectively, in the \textit{Vega system}. We utilize data from the 2MASS ``high-reliability" catalog\footnote{Described at  \url{https://old.ipac.caltech.edu/2mass/releases/allsky/doc/sec2_2.html}.}, which minimizes contamination and confusion caused by neighboring point and/or extended sources. 

\subsection{Gaia DR2} \label{subsec:gaia}

Gaia DR2 is the second data release of the Gaia mission \citep{Gaia_2016}. It provides proper motion and parallax measurements for roughly 1.3 billion stars, along with photometric data in the $G$, $G_{RP}$ and $G_{BP}$ bands. The astrometric catalog we use \citep{Lindegren_2018} has a limiting magnitude for 99.875\% (i.e. $3\sigma$) of sources around $G \approx 21$ mag and a bright limit of $G \approx 3$ mag (see \citet{Brown_2018} for additional details). The typical astrometric uncertainty is around $0.7$ mas for the faintest stars and $0.04$ mas in the bright limit.

In this work, we only incorporate Gaia parallax measurements and their uncertainties. Incorporating Gaia photometric data is deferred to future work.

\subsection{Assembling a Final Catalog} \label{subsec:catalog}

For our northern clouds ($\delta > -30^\circ$), we cross-match all sources in PS1, 2MASS, and Gaia DR2 within a radius of 1 arcsec, with the closest source being selected if there are multiple matches. To limit ourselves to stars for which our distance estimates will likely be most accurate, we adopt the procedure of \citet{Schlafly_2014} and require that the star be detected in the $g_{\rm PS1}$ band as well as at least three of the four other PS1 bands. For our M-dwarf only sightlines towards very nearby clouds (see \S \ref{subsec:selection}), we require that the star be detected in $g_{PS1}$, $r_{PS1}$, and $i_{PS1}$ in order to select a reliable sample of M-dwarf stars. To reduce contamination from galaxies, we require that the aperture magnitude of the star be $< 0.1$ mag brighter than the PSF magnitude in at least three PS1 bands. This is a rather loose cut, but at the Galactic latitudes explored in this work, stars greatly outnumber galaxies in the magnitude range of interest and so we do not expect residual contamination to be a major issue. 

For our southern clouds ($\delta < -30^\circ$), we cross-match all sources in NSC, 2MASS, and Gaia DR2 within a radius of 1 arcsec, with the closest source being selected if there are multiple matches. All of our southern clouds are nearby, so like our northern ``Mdwarf-only" clouds, we require that the star be detected in $g_{\rm DECam}$, $r_{\rm DECam}$, and $i_{\rm DECam}$. Unlike PS1, the NSC catalog does not provide complete coverage in all five optical bands. Thus, we adopt a slightly weaker cut, requiring that the star be detected in at least four bands total, and allow the fourth band to be any DECam or 2MASS band. To reduce contamination from galaxies, we impose the criterion that the NSC's \texttt{class\_star} field be greater than 0.8, indicating that the source has at least an 80\% probability of being a star. 

For both our northern and southern clouds, a Gaia parallax measurement is not required for inclusion into our catalog. However, if a parallax measurement is available for a star, we impose the same quality cuts recommended in \citet{Lindegren_2018} (see their Equation 11). That is, the star must have a mean $G$ magnitude $\leq 21.0$, at least six \texttt{visibility\_periods\_used} and an \texttt{astrometric\_sigma5d\_max} $\leq (1.2 {\rm mas}) \times \gamma(G)$, where $\gamma(G) = {\rm max}[1,10^{0.2(G-18)}]$.

To account for systematic errors in the photometry and possible limitations of our stellar models (see \S\ref{subsec:perstar}), we add a 2\% (0.02 mag) systematic uncertainty in quadrature with the reported errors on the observed magnitudes in all bands. For Gaia parallax measurements, we add in a systematic uncertainty of $0.04$ mas in quadrature with the reported parallax errors.

\section{Methods} \label{sec:method}

Our technique for inferring the clouds distances builds on that of \citet{Schlafly_2014}. Here we summarize our procedure for inferring the per-star distance-extinction estimates and the line-of-sight dust distribution, along with the modifications we make to the \citet{Schlafly_2014} model. 

\subsection{Obtaining Per-Star Distance-Extinction Estimates} \label{subsec:perstar}

\begin{figure}
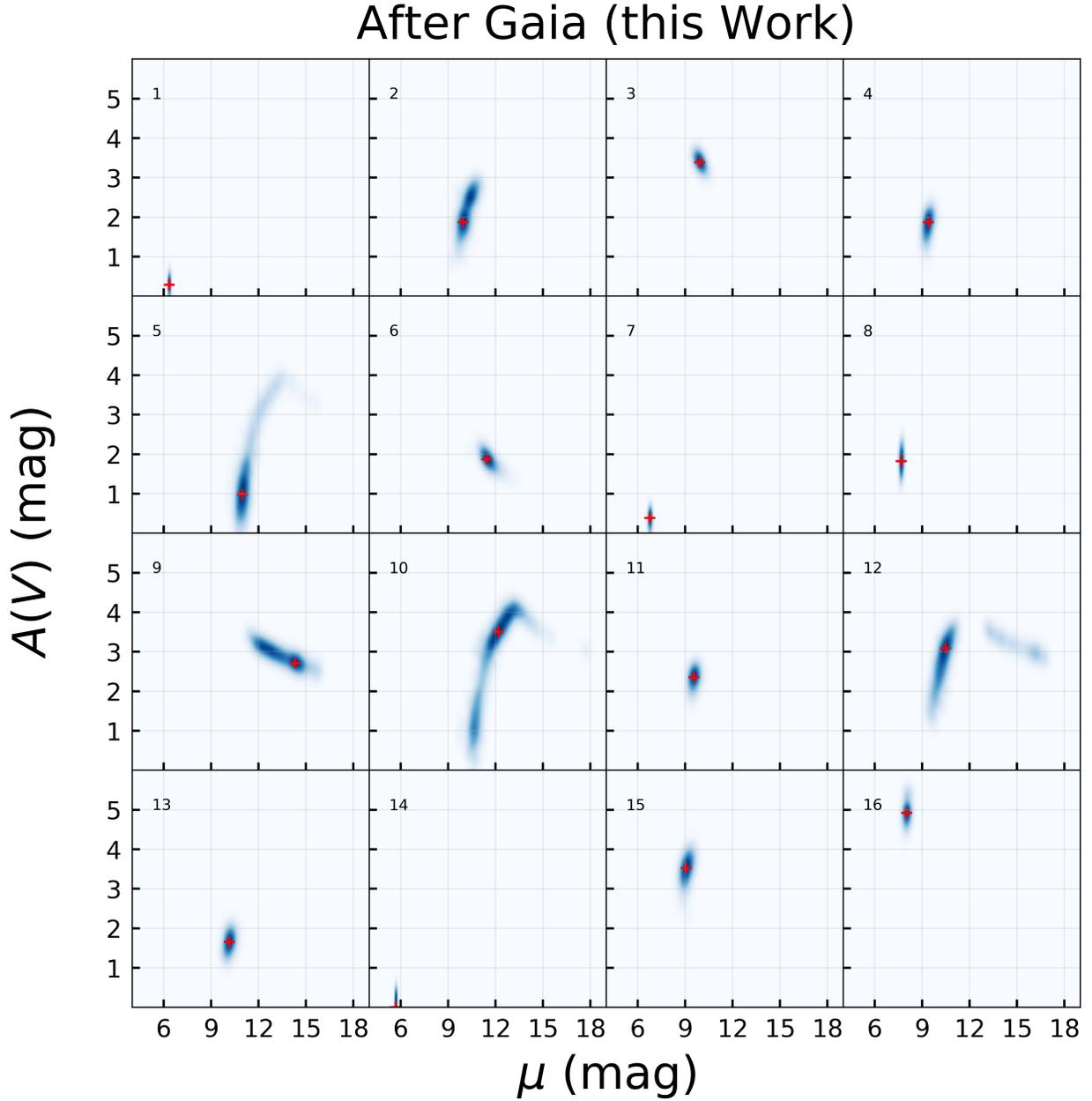

\begin{center}
\includegraphics[width=0.98\columnwidth]{{{stellar_posteriors_newsurfs}}}
\caption{{\label{fig:stellar_posts} A set of 2-D distance-extinction stellar posteriors towards the Perseus molecular cloud ($l=159.3^\circ, b=-20.6^\circ$) after incorporating parallax measurements from Gaia DR2 (see \S \ref{subsec:perstar}). The most probable distance and extinction to each star is marked with a red cross. For high signal-to-noise parallaxes, the Gaia DR2 data alone can significantly constrain the distance modulus (panels 1, 7, 8, and 14). However, even for parallaxes with more moderate signal-to-noise measurements, the Gaia DR2 data help break degeneracies between a star being a nearby ``dwarf'' or a faraway ``giant'' by significantly favoring one solution over the other (panels 3, 9, 11, and 15).
\textbf{An interactive version of this figure showing the differences before and after Gaia is available  \href{https://faun.rc.fas.harvard.edu/czucker/Paper_Figures/stellar_post_comp.html}{\color{blue} here}}.}}
\end{center}
\end{figure}

While our per-star inference pipeline is very similar to those used in past work \citep{Schlafly_2014,Green_2014,Green_2015, Green_2018}, we want to highlight the following changes:
\begin{enumerate}
    \item Unlike Green et al., we measure in units of extinction, $A(V)$, rather than reddening, $E(B-V)$.
    \item We accommodate variations in the extinction curve by allowing $R(V)$ (the optical total-to-selective extinction ratio) to float on a star-by-star basis, substantially improving the flexibility of our stellar modeling.
    \item We incorporate Gaia DR2 parallax measurements as an independent constraint on the distance, tightening the distance constraints to a significant fraction of nearby sources.
    \item We use brute-force methods (rather than Markov Chain Monte Carlo, or MCMC) to derive stellar posteriors, enabling better modeling of distance and extinction estimates for stars displaying significantly extended and/or multi-modal behavior.
\end{enumerate}
These are described in more detail below. \newline

\noindent We model each star as having observed magnitudes $\mathbf{m} = \lbrace m_{g_{\rm PS1}}, \dots, m_{K_{\rm 2MASS}} \rbrace$ as a function of intrinsic stellar type, extinction, and distance, in a manner following \citet{Green_2014, Green_2015, Green_2018}. This gives predicted magnitudes for each source as

\begin{equation}
\mathbf{m} = \mathbf{m}_{\rm int}(M_r, {\rm [Fe/H]}) + A_V(\mathbf{R} + R_V \mathbf{R}') + \mu
\end{equation}

\noindent where $\mu$ is the distance modulus and $\mathbf{m}_{\rm int}$ is the set of intrinsic magnitudes for the star.

Following \citet{Green_2014}, we use an empirical set of models, which we will refer to as the ``Bayestar'' models, calibrated on empirical observations of the stellar locus in low-reddening regions of the sky. These are functions of the absolute $r$-band magnitude \textit{in PS1} ($M_r$) and a vector designed to roughly track the impact of metallicity ([Fe/H]). For PS1 and 2MASS, these models are identical to those used in \citet{Green_2018}. For DECam data from NSC, we have fit a new stellar locus to stars inside the Dark Energy Survey footprint \citep[DES;][]{abbott_18}; see Appendix \ref{sec:decam_models} for details.

The extinction is determined by the integrated amount of dust along the line of sight and the shape of the attenuation curve. The overall amount of dust is parameterized by $A_V \equiv A(V)$, which measures the total attenuation in magnitudes in the $V$ band. The shape of the dust curve is taken to be a function of one parameter $R_V \equiv R(V) \equiv A(V) / E(B-V)$, which measures the attenuation in the $V$ band $A(V)$ relative to the color excess in the $B$ and $V$ bands $E(B-V)$. This serves as a rough proxy for the blueward slope of the attenuation curve, and has been shown to be a reasonable way to parameterize the shape of the attenuation curve in the optical and NIR \citep{Fitzpatrick_1999,Schlafly_2016}. The particular values for $\mathbf{R}$ and $\mathbf{R}'$, which parameterize the mean shape and $R_V$-dependence of the \textit{extinction vector} for our photometry, are derived using the interpolated curve from \citet{Schlafly_2016} over the PS1, DECam, and 2MASS bands assuming the same K-giant model with $T_{\rm eff} = 4500$, ${\rm [Fe/H]} = 0$, and $\log g = 2.5$.\footnote{Technically this curve is not parameterized explicitly as a function of $R(V)$, but the two parameterizations are so similar that the differences are negligible for our purposes.}

Based on our model, the posterior probability that a set of observed magnitudes $\hat{\mathbf{m}}$ is consistent with our predicted photometry $\mathbf{m}(\params) \equiv \mathbf{m}(M_r, {\rm [Fe/H]}, A_V, R_V, \mu)$ and a measured parallax from Gaia DR2 $\hat{\varpi}$ can be computed following Bayes' Rule:

\begin{equation}
P(\params | \hat{\mathbf{m}}, \hat{\varpi}) \propto \mathcal{L}(\hat{\mathbf{m}} | \params) \, \mathcal{L}(\hat{\varpi} | \mu) \, \pi(\params)
\end{equation}

\noindent We assume our likelihood is independent and roughly Gaussian in the measured magnitude in each band $b$ such that
\begin{equation}
\mathcal{L}(\hat{\mathbf{m}}|\params) = \prod_b \frac{1}{\sqrt{2\pi} \sigma_b} \exp\left[{-\frac{1}{2}\frac{(\mathbf{m}(\params)-\hat{\mathbf{m}})^2}{\sigma_b^2}}\right]
\end{equation}
where the product over $b$ is taken over all observed bands. Likewise, we assume that the likelihood for the parallax is Gaussian:
\begin{equation}
\mathcal{L}(\hat{\varpi}|\mu) = \frac{1}{\sqrt{2\pi} \sigma_\varpi} \exp\left[{-\frac{1}{2}\frac{(\varpi(\mu)-\hat{\varpi})^2}{\sigma_\varpi^2}}\right]
\end{equation}

Our prior is comprised of a few separate components that can be factored as

\begin{equation}
\pi(M_r, {\rm [Fe/H]}, A_V, R_V, \mu) = \pi(A_V) \times \pi(R_V) \times \pi(M_r) \times \pi(\mu, {\rm [Fe/H]})
\end{equation}

Since we plan to fit for the extinction along a given sightline (see \S\ref{subsec:los}), we take the prior on $A_V$ to be mostly uninformative and flat between $A_V=0$ mag and $A_V=12$ mag. Our prior for $R_V$ is Normal with a mean of $\langle R_V \rangle = 3.32$ and a standard deviation of $\sigma(R_V)=0.18$ based on observations from \citet{Schlafly_2016}. The prior over $M_r$ is based on measurements from PS1 following \citet{Green_2014}. Finally, the joint prior on distance $\mu$ and metallicity [Fe/H] is derived from a 3-D Galactic model following \citet{Green_2014}, derived from prior knowledge of the number density, metallicity, and luminosity of stars throughout the Galaxy \citep{Ivezic_2008, Juric_2008, Bressan_2012}.

To derive the posteriors for each source, we use the public code \texttt{brutus}\footnote{\href{https://github.com/joshspeagle/brutus}{https://github.com/joshspeagle/brutus}} (Speagle et al. 2019b, in prep.). Similar to the method outlined in \citet{Schlafly_2014}, \texttt{brutus} uses a combination of brute-force and linear optimization methods to compute posteriors over a grid in $M_r$ and [Fe/H]. This avoids convergence and sampling issues associated with Markov Chain Monte Carlo (MCMC) methods used in \citet{Green_2014} and better characterization of extended, multi-modal posteriors. A set of $n=2000$ random samples $\lbrace \params_1, \dots, \params_n \rbrace$ are then drawn from the results and saved. These are then used to marginalize over [Fe/H], $M_r$, and $R_V$ to compute the 2-D posteriors $P(\mu, A_V)$ used in \S\ref{subsec:los}.

The Gaia DR2 astrometric data has two significant effects on our stellar posteriors. These are illustrated in Figure \ref{fig:stellar_posts}, which shows a set of stellar 2-D distance-extinction posteriors for stars towards the Perseus molecular cloud before and after adding in the additional parallax constraints. 

First, for nearby stars, Gaia DR2 parallaxes are often observed at such high signal-to-noise that they are able to constrain the distance modulus $\mu$ to within a few hundredths of a magnitude. Second, even when Gaia DR2 parallaxes are only observed at moderate signal-to-noise, they can still be significantly informative by breaking the degeneracy between nearby ``dwarfs'' and faraway ``giants''. This degeneracy manifests as a bimodality in our stellar posteriors; the inclusion of the Gaia parallaxes is usually able to discriminate between the two modes by suppressing the disfavored mode significantly.

To summarize, our model involves five parameters for each star: the PS1 $r$-band absolute magnitude $M_r$, metallicity [Fe/H], overall extinction $A_V$, attenuation curve shape $R_V$, and distance modulus $\mu$. We derive distance and extinction estimates by comparing the predicted magnitudes $\mathbf{m}(\params)$ and distance modulus $\mu$ to the observed magnitudes $\hat{\mathbf{m}}$ and parallaxes $\hat{\varpi}$ with a set of reasonably-informed priors. Random samples from the posterior are then saved to construct marginalized 2-D posteriors in $\mu$ and $A_V$.

\subsection{Modeling the Line-of-Sight Extinction} \label{subsec:los}

\begin{figure}
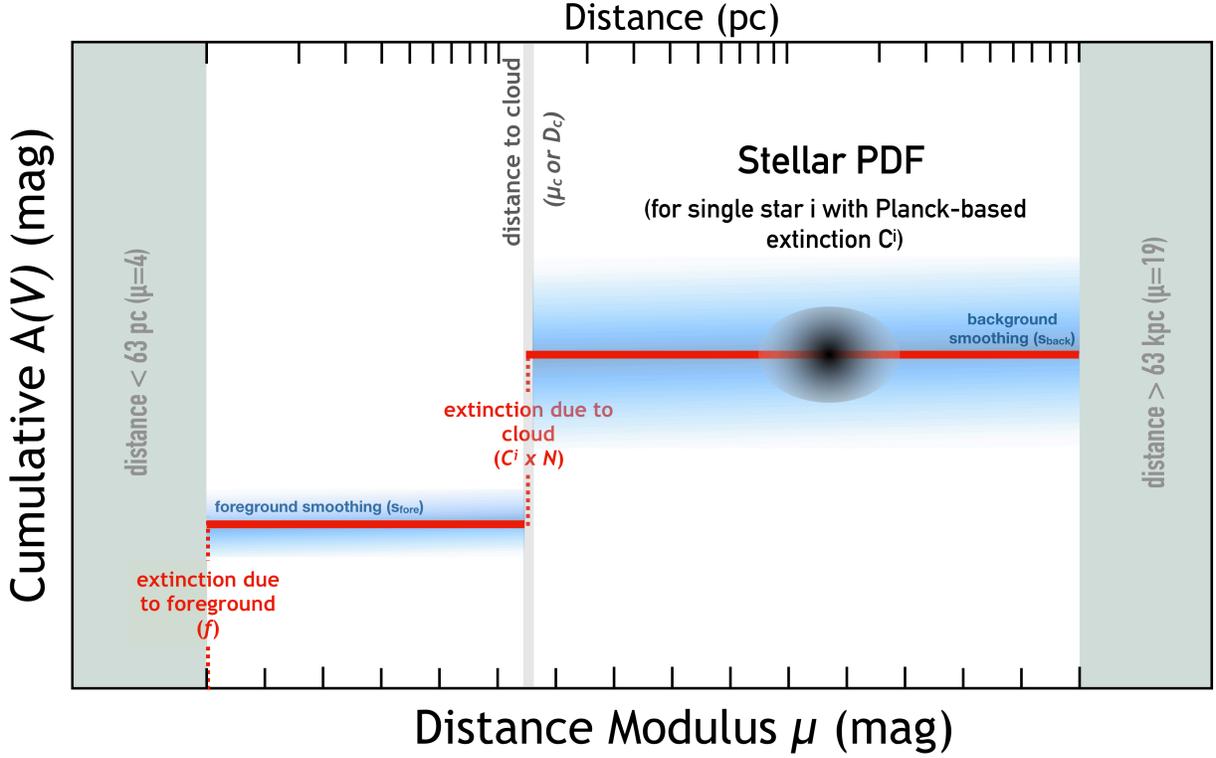

\begin{center}
\includegraphics[width=0.98\columnwidth]{{{schematic_S14}}}
\caption{{\label{fig:example_profile} Cartoon line-of-sight extinction profile demonstrating our basic thin dust-screen model. The mean extinction profile (solid red line) is defined by the free parameters $f$ (foreground cloud extinction), $\mu_c$ or $D_C$ (distance modulus/distance to the molecular cloud), and $N$ (a normalization factor, fixed for all stars, that accounts for any scale difference between the Plank-based extinction and our derived extinction; see \S \ref{subsec:los}). The quantity $C_{i}$ constitutes the Planck-based extinction for star $i$, modeled as the line-of-sight reddening from Planck towards the star times its $R(V)$. Possible extinction variation along the line of sight (shaded blue) is modeled separately before and after the cloud. The resulting extinction profile is overlaid on an idealized 2-D distance-extinction posterior for star $i$ (grayscale ellipsoid); this is akin to the distance-extinction posteriors shown in Figure \ref{fig:stellar_posts}. The likelihood contribution from this star is the integral following the cloud's extinction profile. This star's likelihood would then be multiplied together with the likelihoods of other stars in the sightline to get the total likelihood of the cloud parameters. See \S\ref{subsec:los} for additional details.}}
\end{center}
\end{figure}

\subsubsection{Basic Model} \label{subsubsec:los_basic}
Our basic line-of-sight dust model is very similar to that of \citet{Schlafly_2014}. First, we model the line-of-sight extinction towards each cloud as being dominated by a single thin dust screen at the cloud distance modulus $\mu_C$. The total extinction through the screen towards a single star $i$ in the sightline is parameterized as $N \times C^i$, where $C^i$ is the amplitude of the reddening given by Planck at 353 GHz \citep{Planck_2014} towards star $i$ multiplied by its typical $R_V^i$ (see \S \ref{subsec:perstar}) and $N$ is a normalization factor that corrects for the overall scale difference between the Planck-derived extinction and our per-star ``Bayestar" derived extinction (see \S \ref{subsec:perstar}).\footnote{This is a very simplistic parameterization of the extinction, given that Planck actually measures the long-wavelength emissivity of dust, rather than the extinction in optical bands. The Planck team derives $\tau_{353}$ by modeling the dust emission with a modified blackbody and then converts $\tau_{353}$ to $E(B-V)$ using the correlation between the reddening of quasars and dust optical depth along the same line of sight. We have built in our normalization factor N to account for small errors in our extinction templates, so we expect this will have no effect on our derived distances}

In addition to the bulk of the extinction associated with the cloud, we also account for possible foreground extinction by parameterizing it as a constant $f$. Together these three terms parameterize the line-of-sight extinction profile $A_V(\mu)$ for individual stars through each sightline:

\begin{equation}
\label{eq:profile}
A_V(\mu) = \left\{
        \begin{array}{lr}
            f &  \mu <  \mu_C \\
            f + N \times C^i &  \mu \geq \mu_C \\
        \end{array}
    \right.
\end{equation}

\noindent where, again, $C^i$ is the Planck-based extinction towards each star $i$.

If the extinction profile $A_V(\mu)$ is parameterized by $\cparams = \lbrace \mu_C, f, N \rbrace$ and we had measured the distance modulus $\mu^i$ and extinction $A_V^i$ to each source, then the posterior distribution of $\cparams$ would be given by:

\begin{equation}
P(\cparams | \lbrace \mu^i, A_V^i \rbrace ) \propto \mathcal{L}(\lbrace \mu^i, A_V^i \rbrace | \cparams) \, \pi(\cparams) = \pi(\cparams) \times \prod_i \mathcal{L}( \mu^i, A_V^i | \cparams)
\end{equation}

\noindent assuming our measurements for each star $i$ are independent and some prior $\pi(\cparams)$ over the cloud parameters. Unfortunately, we do not have an exact measurement of $\mu^i$ and $A_V^i$ for each star, but rather a posterior estimate $P(\mu, A_V | \hat{\mathbf{m}}^i, \hat{\varpi}^i)$. We thus do not know precisely what distance or extinction that star actually has, complicating this comparison.

Ideally, we would like to model the joint distribution of our line-of-sight parameters $\cparams$ along with the individual parameters $\params^i$ for each star. We defer this to future work. Instead, we will simply integrate over each per-star posterior to try and marginalize over this uncertainty. Our likelihood $\mathcal{L}( \mu^i, A_V^i | \cparams)$ then becomes:

\begin{equation}
\label{eq:like}
\mathcal{L}(\cparams | \hat{\mathbf{m}}_i) = \int P(\cparams|\mu, A_V) \, P(\mu, A_V | \hat{\mathbf{m}}_i) \, d\mu \, dA_V
\end{equation}

\noindent where $P(\mu, A_V | \hat{\mathbf{m}}_i)$ is the per-star posterior and $P(\cparams|\mu, A_V)$ implements the line-of-sight model for $A_V(\mu)$ outlined above:

\begin{equation}
P(\cparams | A_V, \mu)  = \left\{
        \begin{array}{lr}
            \delta(f) &  \mu <  \mu_C \\
            \delta(f + N \times C^i) &  \mu \geq \mu_C \\
        \end{array}
    \right.
\end{equation}

\noindent where $\delta(\cdot)$ is the Dirac delta function.

To summarize, our basic model assumes that a given cloud is at some distance modulus $\mu_C$ and functions as a thin dust screen with amplitude $N$ relative to Planck. In front of each cloud is a small amount of foreground extinction $f$. Because we do not know the exact distance and extinction of each star, we marginalize over them by integrating over the 2-D distance-extinction posterior for each star (from \S\ref{subsec:perstar}) relative to our line-of-sight model.

\subsubsection{Modifications} \label{subsubsec:los_mod}

We build on the basic formalism described above in three main ways:
\begin{enumerate}
    \item We allow some dispersion in the foreground extinction to account for possible variation in a given spatial region.
    \item We allow for variation within each cloud (i.e. the ``background'') to account for patchiness in cloud geometry relative to the Planck-based dust screen model.
    \item We account for possible outliers in the stars used to model the cloud that can arise due to mis-estimated posteriors or more complex 3-D dust geometry.
\end{enumerate}

We describe these each in turn. See Figure \ref{fig:example_profile} for a schematic illustration of this model. \newline

\noindent As discussed in \S\ref{subsubsec:los_basic}, our basic model assumes that there is only a single possible extinction for the foreground or the background. However, it is likely that there is some variation within a given spatial region. We account for this by assuming our model actually has some dispersion such that the probability becomes:

\begin{equation}
\label{eq:profile2}
P(\cparams | A_V, \mu)  = \left\{
        \begin{array}{lr}
            \mathcal{N}(f, \sigma_{\rm fore}^2) &  \mu <  \mu_C \\
            \mathcal{N}(f + N \times C^i, \sigma_{\rm back}^2) &  \mu \geq \mu_C \\
        \end{array}
    \right.
\end{equation}

\noindent where $\mathcal{N}(\cdot,\cdot)$ indicates a normal distribution with the listed mean and variance, respectively. Here, we define the relevant standard deviations with respect to a ``smoothing parameter'' $s$, which is parameterized as a \textit{fraction} of the total $A_V$ range explored by the model, i.e.
\begin{equation}
s_{\rm fore} = \frac{1}{A_V^{\rm max} - A_V^{\rm min}} \times \sigma_{\rm fore}
\end{equation}

\noindent and likewise for $s_{\rm back}$. 

Alternately, we can interpret $s_{\rm fore}$ and $s_{\rm back}$ as adaptively smoothing the per-star posteriors along the extinction axis with respect to a given model $\cparams$ for the cloud. The portion of the star's 2-D distance-extinction posterior in front of the cloud ($\mu < \mu_C$) is smoothed with a Gaussian kernel with $\sigma_{\rm fore}$, while the portion behind the cloud ($\mu \geq \mu_C$) is smoothed with a Gaussian kernel with $\sigma_{\rm back}$.

It is also necessary to consider the effect of outlying stars on this likelihood. \citet{Schlafly_2014} reduced the effect of outliers by adding a small \textit{fixed} constant to every stellar posterior, where the constant was chosen so that some fraction of stars in the sightline was considered to be drawn from a flat outlier distribution in $\mu$ and $E(B-V)$. We adopt a similar approach, except we explicitly seek to model this term.

Specifically, we introduce a final parameter $P_b$, which models our individual likelihoods as a \textit{mixture model} following \citet{Hogg_2010}. This model assumes that a given star $i$ is associated with our given cloud model $\cparams$ with probability $1-P_b$, and is associated with a particular outlier model with probability $P_b$. Broadly speaking, it quantifies the fraction of ``bad'' stars that are inconsistent with our model.

Our mixture model modifies our likelihood to be:

\begin{equation}
\label{eq:outlier}
\mathcal{L}(\cparams | \hat{\mathbf{m}}_i) = (1 - P_b) \, \mathcal{L}_C(\cparams | \hat{\mathbf{m}}_i) +  P_b \, \mathcal{L}_b(\cparams)
\end{equation}

\noindent where $\mathcal{L}_C(\cparams | \hat{\mathbf{m}}_i)$ is the original cloud-based likelihood defined in Equation \ref{eq:like} and $\mathcal{L}_b(\cparams)$ is the likelihood under our assumed outlier model. We take this to be uniform over distance modulus $\mu$ and extinction $A_V$ within the bounds of our priors, which gives a constant value of

\begin{equation}
\mathcal{L}_b(\cparams) = \mathcal{L}_b = \frac{1}{A_V^{\rm max} - A_V^{\rm min}}
\end{equation}

\subsection{Priors} \label{subsec:los_priors}

Our Bayesian approach requires that we adopt priors on our model parameters $\cparams$ for the cloud. These now include six parameters:
\begin{itemize}
    \item the cloud distance modulus ($\mu_C$), 
    \item the foreground extinction ($f$),
    \item the normalization factor ($N$), 
    \item the foreground smoothing ($s_{\rm fore}$),
    \item the background smoothing ($s_{\rm back}$), and
    \item the outlier fraction ($P_b$).
\end{itemize}

We impose a flat prior on both the cloud distance modulus $\mu_C$ and the normalization factor $N$, such that $4 < \mu_C < 14$ mag and $0.2 < N < 2$. We also impose a conditional flat prior on the foreground extinction, requiring that it be less than 25\% of the median projected extinction towards the stars determined by the Planck extinction and each star's $R_V$. We impose a truncated Log-Normal prior over the range 0 to 1 on both smoothing parameters $s_{\rm fore}$ and $s_{\rm back}$, with a mean of 0.05 and a standard deviation equivalent to a factor of 1.35. Finally, we impose a truncated Log-Normal prior on our $P_b$ parameter over the range 0 to 1, with a mean of 0.03 and a standard deviation equivalent to a factor of 1.35.

\input{averagedist_table.tex}

\subsection{Sampling} \label{subsec:los_sample}

We sample for our set of model parameters ($\mu_C$, $f$, $N$, $s_{\rm fore}$, $s_{\rm back}$, $P_b$) using the public nested sampling code \texttt{dynesty}\footnote{\href{https://github.com/joshspeagle/dynesty}{https://github.com/joshspeagle/dynesty}} \citep{Speagle_2019}. The cloud distance $D(\mu_C)$ is our primary free parameter of interest; we consider all other parameters, as well as the individual stellar posteriors, as nuisance parameters to be marginalized over, although we report them for completeness. See Appendix \ref{sec:dynesty} for more information on the sampling routine we implement.

\subsection{Sample Selection} \label{subsec:selection}

Because the target clouds lie at a range of distances (from $< 100$ pc to $> 2$ kpc), we adopt two slightly different techniques for distance determination following \citet{Schlafly_2014}. For clouds which are ``far away'' ($D_C \gtrsim 200$ pc), stars of all stellar types are used. For clouds which are ``nearby'' ($D_C \lesssim 200$ pc, covering Aquila S., Ophiuchus, Taurus, Hercules, and all the MBM clouds with the exception of MBM 46-47), only M-dwarf-like stars are used. Limiting our analysis to only M-dwarfs for very nearby clouds prevents the small number of foreground stars from being overwhelmed by the large number of background stars when trying to pinpoint where a step in extinction occurs (see \S\ref{subsec:los}).

We select for M-dwarfs by imposing the following cuts based on  \citet{Schlafly_2014}:

\begin{equation}
\label{eq:brightred}
g - \frac{A_g}{A_g - A_r} (g - r - 1.2) < 20 \\
\end{equation}
\begin{equation}
\label{eq:colred}
r - i - \frac{A_r - A_i}{A_g - A_r} (g - r - 1.2) > 0.78
\end{equation}

\noindent where the extinction coefficients $A_x$ are taken from Table 1 of \citet{Green_2018}. \noindent Both of these equations impose color and magnitude cuts along the reddening vector. Equation \ref{eq:brightred} primarily selects bright blue stars (as faint as $g_{P1}$=20 mag), which have a typical unreddened $g_{P1} - r_{P1}=1.2$. Equation \ref{eq:colred} selects M-dwarfs in color-color space. As argued in \citet{Schlafly_2014}, the application of both cuts produces a clean, reddening-independent sample of nearby M-dwarfs ideal for the ``near" analysis. Compared to \citet{Schlafly_2014} we make one minor revision to Equation \ref{eq:colred}, by fixing the intercept of the cut to 0.78; this ensures that we intersect the $g_{P1}-r_{P1}$ vs. $r_{P1}-i_{P1}$ color-color diagram at the same location as \citet{Schlafly_2014} for a $g_{P1}-r_{P1}=1.2$, despite the adoption of the updated reddening vector. 

In addition to the above cuts, we also follow \citet{Schlafly_2014} by masking out all stars with Plank-based $E(B-V) < 0.15$ mag, independent of $R(V)$. Similar to the M-dwarf cut, limiting our sample to stars with at least $0.15$ mag of reddening (assuming they are in the background) ensures that only stars capable of informing where a jump in reddening occurs are used in the fit.

Finally, we remove stars from our analysis with poor best-fit chi-square values $\chi^2_{\rm best}$, which indicate these stars are likely inconsistent with our stellar models and therefore likely have unreliable distance-extinction estimates. Since the expected $\chi^2$ for any particular source depends on the number of bands observed, we opt to remove all stars with

\begin{equation}
P(\chi^2_{n_{bands}} > \chi^2_{\rm best}) < 0.01
\end{equation}

\noindent where $n_{bands}$ is the number of bands observed. This is roughly equivalent to removing outliers at approximately the $2\sigma-3\sigma$ level. We note that this chi-squared cut will likely remove many of the stars embedded in the clouds themselves. This is because most of the embedded stars are likely to be young, and our stellar templates only robustly model older, main sequence stars. Since our method leverages bracketing the cloud between foreground and background stars in the context of a thin dust screen model, we expect any young stars embedded in the clouds themselves (with unreliable distance-extinction measurements) will have a negligible effect on our results.

\subsection{Uncertainties} \label{subsec:errors}

Our Bayesian approach enables a robust determination of \textit{statistical} uncertainties through our posterior samples (see \S\ref{subsec:los_sample}). However, there are additional \textit{systematic} uncertainties in our approach that need to be considered. These stem from three main factors:
\begin{enumerate}
    \item \textit{Errors in the stellar models}: Our empirical stellar models are derived from fits to the observed stellar locus in PS1, 2MASS, and NSC DECam data. These likely contain some amount of residual noise, possible residual reddening, and extrapolated metallicity dependence. Slight differences between our models and the intrinsic colors of stars could result in systematic extinction/distance offsets for each star.
    \item \textit{Errors in the extinction vector:} Although we have based our extinction vector and its variation on empirical results from \citet{Schlafly_2016}, it is possible that the mean vector and/or $R(V)$ dependence is slightly off. This mismatch could result in systematic extinction/distance offsets for each star.
    \item \textit{Errors in the line-of-sight cloud model:} Our cloud model (see \S\ref{subsec:los}) is quite simplistic, assuming that the majority of the cloud is located at a single distance (modulus) $\mu_C$, that any foreground stars only have a single, uniform extinction $f$, and that neighboring regions of the sky are completely independent from each other. Any more complex behavior that violates these assumptions can systematically bias $\mu_C$.
\end{enumerate}

Compared to the analysis presented in \citet{Schlafly_2014}, we are much less impacted by systematic uncertainties arising from our per-star modeling. All of the stellar models, extinction vectors, and $R(V)$ dependencies used here are significantly improved from that work, and have been tested rigorously in other detailed analyses \citep{Schlafly_2016,Green_2015,Green_2018}. In addition, in many cases the Gaia DR2 parallaxes significantly constrain the distance to nearby stars, further limiting the impact of systematic offsets. Internal testing based on Gaia DR2 data lead us to estimate the impact of any systematics arising from our per-star modeling to be $\lesssim 2\%$.

Instead, the dominant source of systematic uncertainty stems from the simplicity of our line-of-sight dust model. Based on our results in \S\ref{subsec:averages}, we find that these assumptions lead to $5\%$ systematic uncertainty in distance, which we adopt for all our reported values. Since the characteristic statistical uncertainty on the sightlines is often $2-3\%$, the total estimated uncertainty is often dominated by this term.

\begin{figure}[h!]
\begin{center}
\includegraphics[width=0.5\columnwidth]{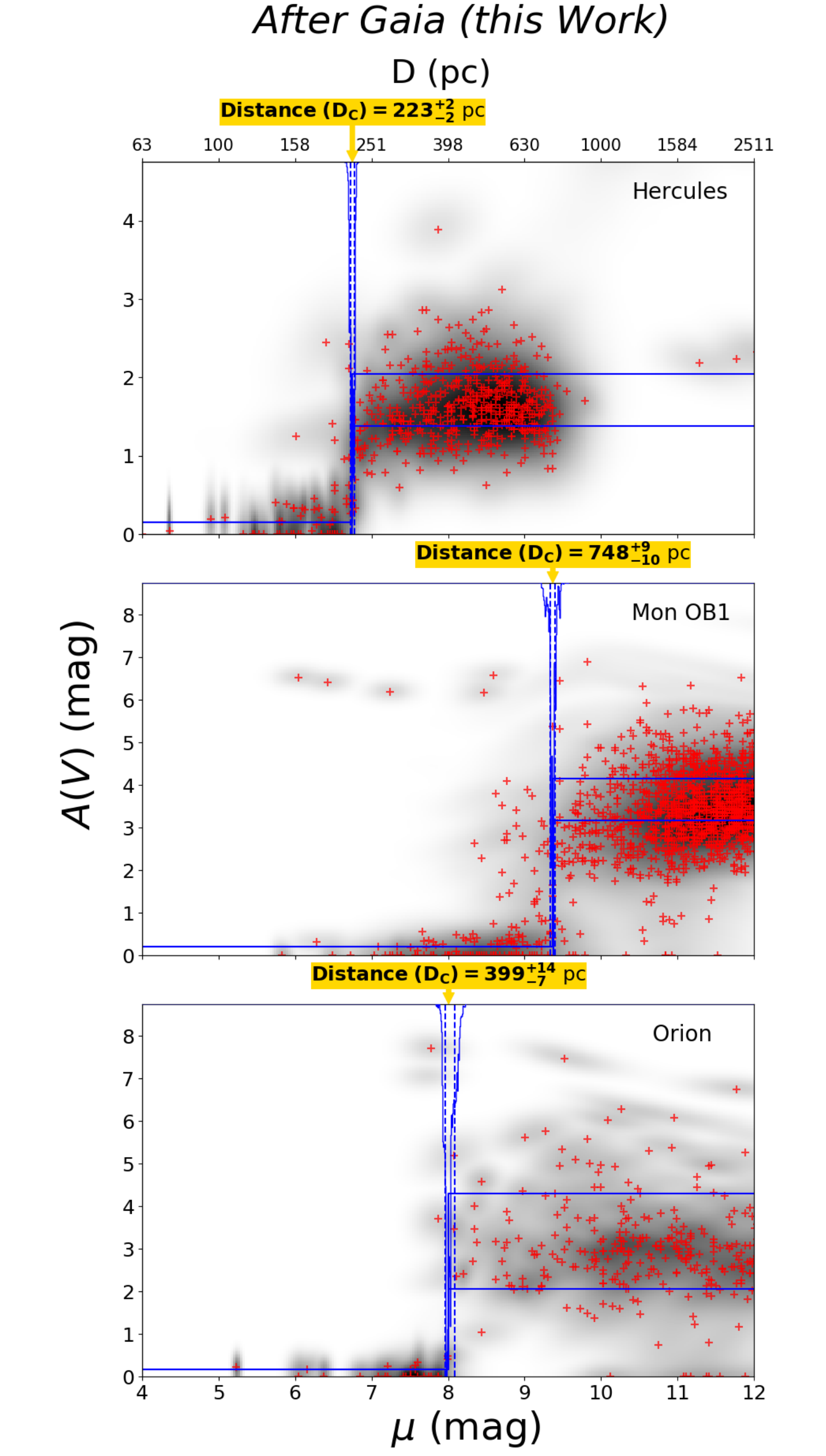}
\caption{ \label{fig:los} Illustration of our fitting procedure for three sightlines from \citet{Schlafly_2014}, with a sightline from Hercules ($l=45.1^\circ, b=8.9 ^\circ$) shown in the top panel, Mon OB1 ($l=201.4^\circ, b=1.1 ^\circ$) in the middle panel, and Orion ($l=208.4^\circ, b=-19.6 ^\circ$) in the bottom panel. The background grayscale shows the stacked distance-extinction posteriors for all the stars used in the fit. The most probable distance and extinction to every star is marked via a red cross. The blue line shows the typical extinction profile inferred for the sightline. Prior to the cloud distance $D_C$ (at distance modulus $\mu_C$), the magnitude of the extinction profile is equal to the foreground extinction, with the median foreground extinction $f$ shown via the blue horizontal line. The probable range of distances to the cloud is shown via the inverted blue histogram at the top of each panel, with the median cloud distance marked via the blue vertical line. On either side we show the 16th and 84th percentile of the cloud distance (derived from our \texttt{dynesty} ``chain") via the vertical blue dashed lines. Beyond the cloud distance, the bottom and top blue lines show the 16th and 84th percentile of the Planck-based extinction towards the stars, multiplied by the median normalization factor $N$ inferred for the sightline. The distance uncertainties do not include any systematic uncertainties, which we estimate to be $5\%$ for this work and 10\% for \citet{Schlafly_2014}. \textbf{An interactive version of this figure showing the differences before and after Gaia is available \href{https://faun.rc.fas.harvard.edu/czucker/Paper_Figures/threepanel_comp.html}{\color{blue} here}.}}
\end{center}
\end{figure}

\section{Results} \label{sec:results}

We present two separate collections of distances to regions of local molecular clouds. For consistency with past work, in \S\ref{subsec:sightlines} we first apply our analysis to the same sightlines used in \citet{Schlafly_2014} to determine an updated catalog of distances. After investigating any relevant changes, in \S\ref{subsec:pixels} we instead apply our analysis over a larger area of the cloud by systematically subdividing each cloud using a pixelization scheme and independently fitting distances to each pixel. This second technique is then applied to derive 2-D spatial maps to many molecular clouds in the \citet{Dame_2001} CO survey. These results are then used to compute robust average distances to these regions in \S\ref{subsec:averages}, which are reported in Table \ref{tab:avgtab}.

\subsection{\citet{Schlafly_2014} Sightlines} \label{subsec:sightlines}

\citet{Schlafly_2014} presented a catalog of distances motivated by a combination of the \citet{Magnani_1985} (MBM) molecular cloud catalog and the \citet{Dame_2001} CO survey. Clouds in the MBM catalog were associated with an explicit list of sightlines from that work, and \citet{Schlafly_2014} adopted the same sightlines to remain consistent. Clouds from the \citet{Dame_2001} CO survey, however, did not have an explicit set of sightlines associated with them. As a result, \citet{Schlafly_2014} instead chose a representative set of sightlines that was well-suited to the basic method outlined in \S\ref{sec:method} but not intended to provide complete areal coverage of these clouds. 

In total, \citet{Schlafly_2014} included all clouds in the \citet{Dame_2001} CO survey and \citet{Magnani_1985} catalog which fell inside the PS1 footprint ($\delta > -30^\circ$), possessed an adequate number of stars with Planck-based $E(B-V) > 0.15$ mag, and were far enough from the Galactic plane to avoid possible multi-cloud confusion. In this work, we provide distance tables to the \citet{Schlafly_2014} sightlines for completeness in Appendix \ref{sec:sightlines_2014}, but refer readers to \S \ref{subsec:pixels} and \ref{subsec:averages} for the main data products of this paper. 

In Figure \ref{fig:los}, we show the results of our fitting procedure for three different sightlines, in Hercules ($l=45.1^\circ, b=8.9 ^\circ$), Mon OB1 ($l=201.4^\circ, b=1.1 ^\circ$), and Orion ($l=208.4^\circ, b=-19.6 ^\circ$). Although these span a wide range in distance and extinction, we find our model provides an excellent fit to the data, with a well-constrained distance modulus $\mu_C$, small outlier fraction $P_b$, and low foreground extinction $f$ for each sightline. As expected, the majority of the distance constraint comes from a collection of highly-extincted stars \textit{at the same distance}, with the outlier model suppressing the impact of isolated foreground stars with anomalously high extinction. We also find that our model prefers significant amounts of variation in background star extinction (via $s_{\rm back}$) but not the foreground (via $s_{\rm fore}$), illustrating the importance of explicitly modeling this intra-cloud variation.

\begin{figure}[t!]
\begin{center}
\includegraphics[width=1.0\columnwidth]{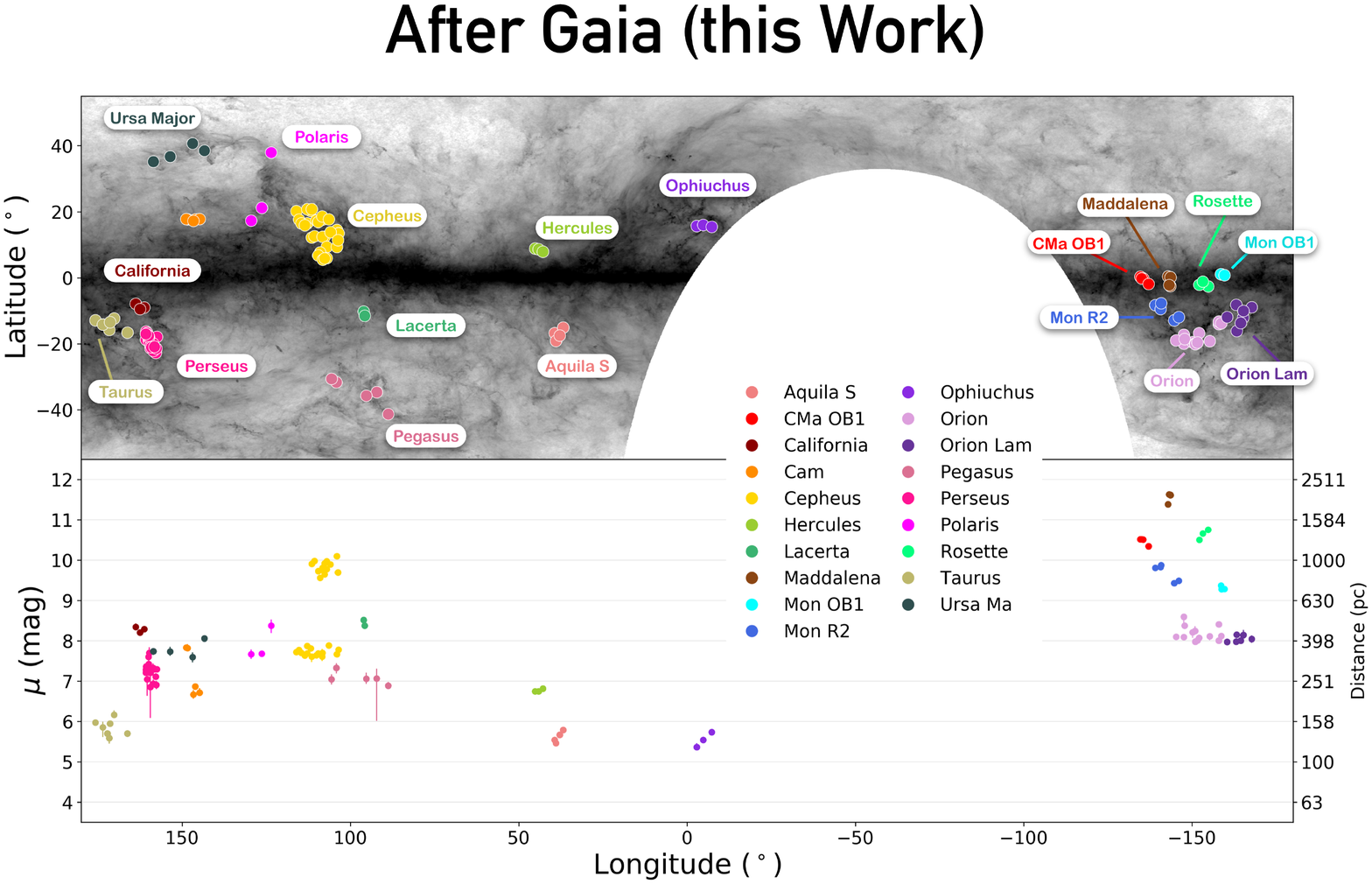}
\caption{ \label{fig:major_summary} Summary of cloud distance results for sightlines through major molecular clouds from the \citet{Dame_2001} CO survey. The top panel shows the Planck reddening map \citep{Planck_2014}, with the central ($l$,$b$) position of each sightline overlaid. The bottom panel shows the inferred distance to the cloud as a function of Galactic longitude, along with estimated statistical uncertainties; systematic uncertainties are excluded, which we estimate to be $5\%$ for this work and $10\%$ for \citet{Schlafly_2014}. Sightlines through the same molecular cloud are grouped by color and labeled by name. \textbf{An interactive version of this figure showing the differences before and after Gaia is available \href{https://faun.rc.fas.harvard.edu/czucker/Paper_Figures/bigcloud_html_final.html}{\color{blue} here}.}}
\end{center}
\end{figure}

\begin{figure}[t!]
\begin{center}
\includegraphics[width=1.0\columnwidth]{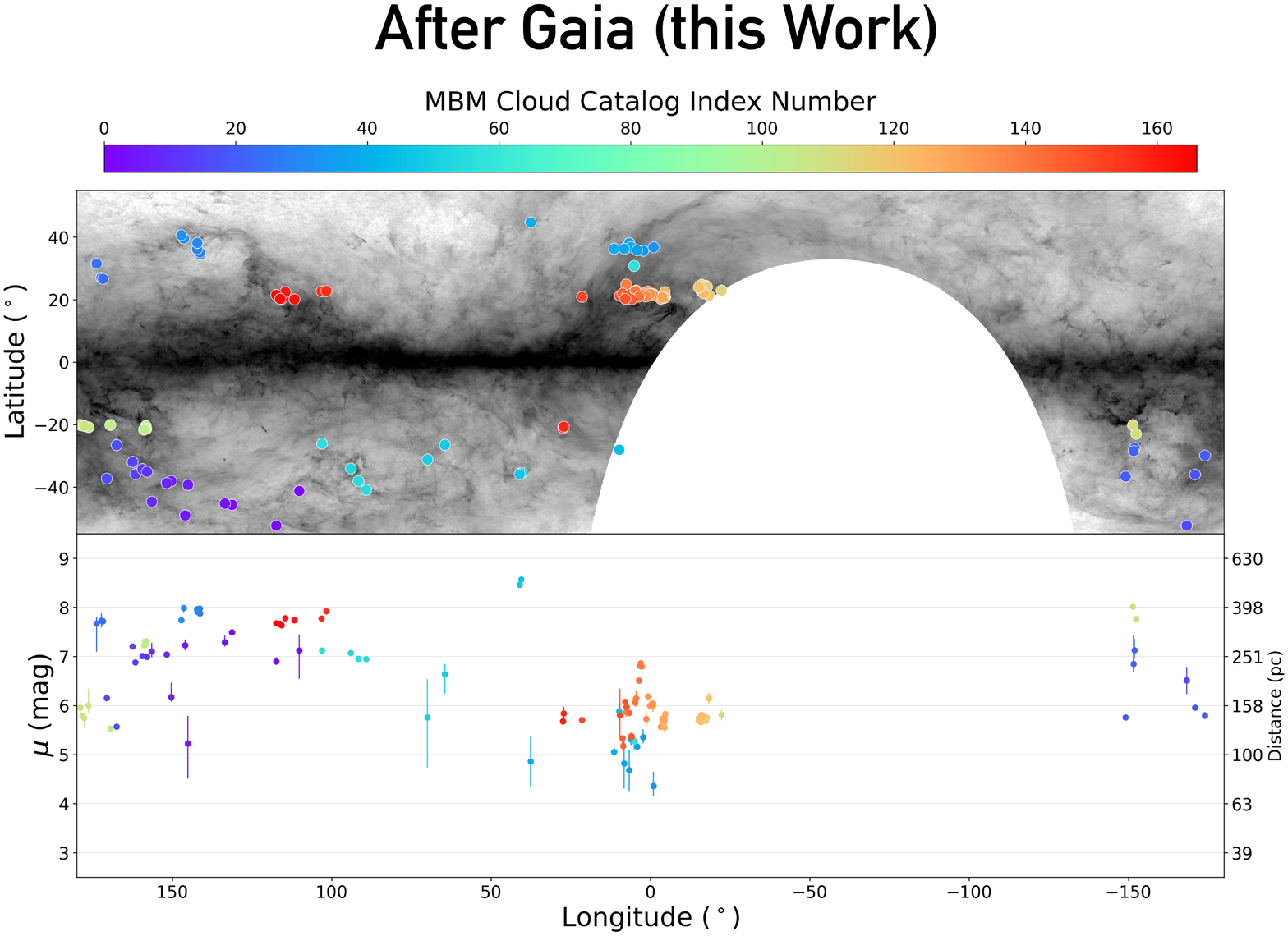}
\caption{ \label{fig:mbm_summary} Same as in Figure \ref{fig:major_summary} --- except now showing the MBM clouds --- with sightlines color-coded according to their cloud index from \citet{Magnani_1985}. \textbf{An interactive version of this figure showing the differences before and after Gaia is available \href{https://faun.rc.fas.harvard.edu/czucker/Paper_Figures/mbmcloud_html_final.html}{\color{blue} here}.}}
\end{center}
\end{figure}

In Figures \ref{fig:major_summary} and \ref{fig:mbm_summary}, we visualize the distance results for every sightline from \citet{Schlafly_2014} (see \S \ref{sec:sightlines_2014} in the Appendix), which includes many major molecular clouds from the \citet{Dame_2001} CO survey and \citet{Magnani_1985}, respectively. Overall, we see that the statistical uncertainties decreases dramatically for most sightlines compared to \citet{Schlafly_2014}, with the dispersion between many clouds ``tightening'' substantially, as expected if they are part of a localized structure at a single distance.

\begin{figure}[t!]
\begin{center}
\includegraphics[width=0.75\columnwidth]{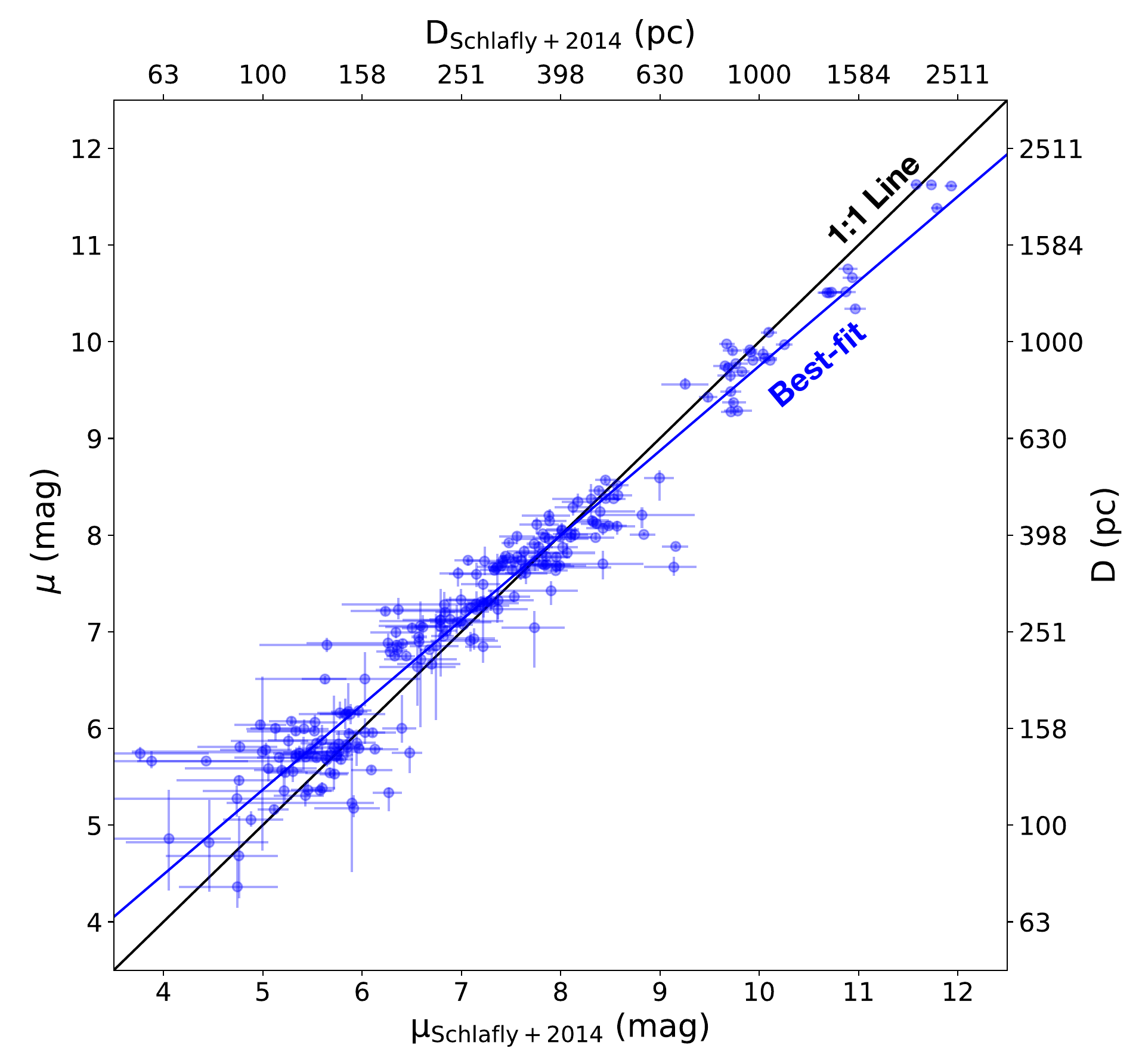}
\caption{ \label{fig:static_comp} A comparison of the distances obtained in \citet{Schlafly_2014} (x-axis) versus distances obtained post-Gaia in this work (y-axis), over the same set of sightlines summarized in Tables \ref{tab:major_tab} and \ref{tab:mbm_tab}. We show the 1:1 line in black and the best-fit line to the distribution in blue (parameterized as $y=0.88(\pm 0.02)x + 0.98(\pm 0.11)$). The uncertainties shown only include the statistical uncertainties, and not any systematic uncertainty, which we estimate to be $\approx 10\%$ for \citet{Schlafly_2014} and $\approx 5\%$ for this work.}
\end{center}
\end{figure}

One interesting feature we observe is that the distances appear to systematically shift depending on a sightline's position relative to the galactic center, with most of the distances to the right of the galactic center (which are also further away) decreasing while those on the left (which are also more nearby) increasing. In both cases, the fractional difference between the \citet{Schlafly_2014} distances and our new Gaia-informed distances is $\lesssim 15\%$, which was the magnitude of the systematic uncertainty estimated in \citet{Schlafly_2014}. This effect is seen more clearly in Figure \ref{fig:static_comp}, where we plot the distances to all \citet{Schlafly_2014} sightlines (based on both \citet{Dame_2001} and \citet{Magnani_1985}) on the x-axis, versus the distances we obtain to the same sightlines post-Gaia on the y-axis. The errorbars shown only include the statistical uncertainties (computed the same way for both \citet{Schlafly_2014} and this work) and do not include any systematic uncertainty, which is $\approx 10\%$ for \citet{Schlafly_2014} and $\approx 5\%$ for this work. As seen in Figure \ref{fig:static_comp} we tend to infer closer distances than \citet{Schlafly_2014} for sightlines with $\mu > 9$ mag, while we tend to infer farther distances with respect to \citet{Schlafly_2014} for sightlines with $\mu < 9$ mag. When modeling the statistical errors on both x and y, the best-fit line is parameterized by: $y=0.88(\pm 0.02)x + 0.98(\pm 0.11)$, as computed using orthogonal distance regression.

\subsection{Detailed Maps of Cloud Structure} \label{subsec:pixels}

To complement the \citet{Schlafly_2014} technique, we also adopt an alternative approach to provide more uniform detailed maps of cloud structure. Guided by previous studies of these clouds in the literature, we start by defining rectangular boundaries (in Galactic longitude and latitude) around each region (see Table \ref{tab:avgtab}). We then subdivide each area into \texttt{healpix} pixels \citep{Healpix_2004} of either \texttt{nside=64} or \texttt{nside=128}, corresponding to pixel areas of 0.84 deg$^2$ and 0.21 deg$^2$, respectively. The larger \texttt{nside=64} pixels are used for all ``nearby'' M-dwarf only sightlines, while the smaller \texttt{nside=128} pixels are used for all ``faraway'' sightlines (see \S\ref{subsec:selection}).

After replacing the circular beams from the sightlines presented in \S\ref{subsec:sightlines} with these \texttt{healpix} pixels, we need to ensure that the only pixels included in our modeling are those that actually trace the cloud, rather than a lower density component along the line-of-sight. To do this, we impose Planck $E(B-V)$ thresholds on the pixels for each cloud and only consider pixels with appreciable amounts of extinction. These thresholds, outlined in Table \ref{tab:avgtab}, vary from 0.15 mag for very diffuse clouds at high latitudes (e.g. Ursa Major) to 3 mag for clouds in the Galactic plane and oriented towards the Galactic center (e.g. Serpens/AqR). 

Once we have a set of pixels within each longitude-latitude box that satisfy this $E(B-V)$ threshold, we apply the exact same procedure as before to estimate the distances to extinction jumps within each pixel.\footnote{The one exception is we limit the maximum number of stars per pixel to 2000 before we apply our chi-squared cut for computational expediency.} This procedure then results in a grid of distance estimates to different parts of each cloud. We are able to provide complete coverage above each cloud's reddening threshold for a vast majority of clouds. The exceptions are the southern clouds Chamaeleon and Lupus, whose pixel coverage is sparse due to the lack of available broadband photometry in the $g$, $r$, and $i$ bands. For Lupus, while we targeted pixels over the entire longitude range $335^\circ < l < 348 ^\circ$, only pixels with $l > 338^\circ$ had enough stars with multi-band optical photometry to reliably fit a distance. Thus, our average distance for Lupus is drawn from pixels in the vicinity of Lupus 1, 3, 5, 6, and 9, and excludes 2, 4, 7, and 8. For Chamaeleon, we provide complete coverage of Chamaeleon I and II, but no coverage of Chamaeleon III. Finally, in a very small fraction of cases ($\approx 1\%$ of sightlines), we have removed pixels from our maps which contain no foreground stars, leading to very uncertain distance estimates. 

\begin{figure}[t!]
\begin{center}
\includegraphics[width=1.0\columnwidth]{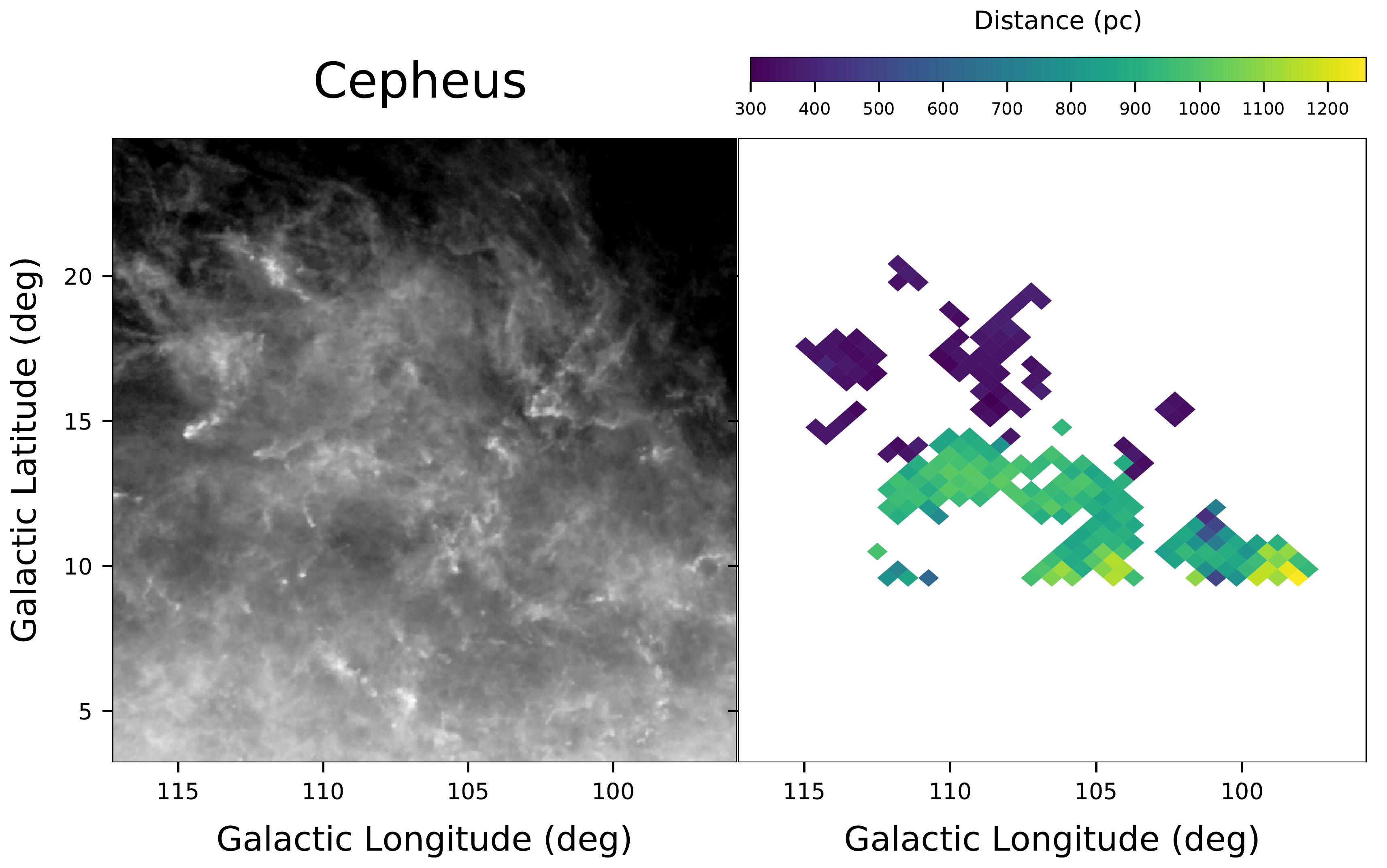}
\caption{ \label{fig:cepheus_hp} Pixelated distance map for the Cepheus molecular cloud. The left panel shows the Planck $E(B-V)$ reddening map in the region around Cepheus. The right shows the set of \texttt{healpix} pixels associated with the cloud (see Table \ref{tab:avgtab} and \S\ref{subsec:pixels}) colored by their median distance. We find Cepheus clearly hosts two separate components: a ``near'' component (purple pixels) and a ``far'' component (yellow or green pixels). \textbf{An interactive version of this figure that includes each pixel's line-of-sight extinction profile is available \href{https://faun.rc.fas.harvard.edu/czucker/Paper_Figures/Cepheus.html}{\color{blue}here}}. Similar figures for every cloud in Table \ref{tab:avgtab} are accessible via Figure \ref{fig:summary}. A machine readable table of the per pixel results for every pixel shown above in Cepheus is available on the \href{https://dataverse.harvard.edu/dataverse/cloud_distances}{Dataverse} (doi:10.7910/DVN/74Y5KU). Similar tables are available for every cloud in Table \ref{tab:avgtab}.}
\end{center}
\end{figure}

\begin{figure}[h!]
\begin{center}
\includegraphics[width=21cm,angle=90]{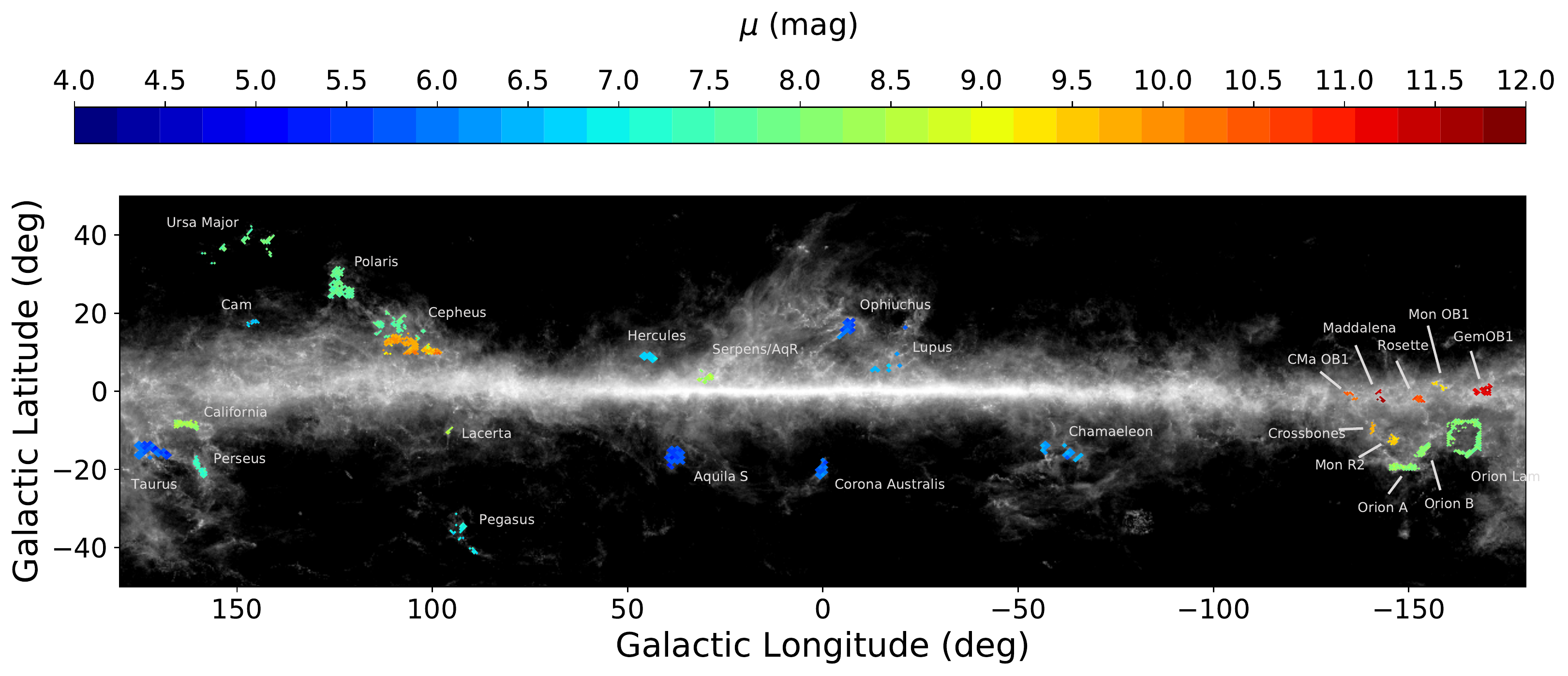}
\caption{ \label{fig:summary} Distance estimates to local molecular clouds through $\approx 1000$ pixelated lines of sight. Each pixel is color-coded according to its median distance modulus and overlaid on the Planck $E(B-V)$ reddening map. See \S\ref{sec:method} for details on the method and \S\ref{subsec:pixels} for details on how the pixels were selected. \textbf{An interactive version of the figure that highlights more detailed results for each cloud akin to Figure \ref{fig:cepheus_hp} is available \href{https://faun.rc.fas.harvard.edu/czucker/Paper_Figures/summary_fig.html}{\color{blue}here}.}}
\end{center}
\end{figure}

An example of the distance map produced for a particular cloud (Cepheus) is shown in Figure \ref{fig:cepheus_hp}. A summary of our results across all pixels is shown in Figure \ref{fig:summary}. Detailed results (including coordinates, distances, etc. for each pixel) for every cloud are available online on the \href{https://dataverse.harvard.edu/dataverse/cloud_distances}{Dataverse} (doi:10.7910/DVN/74Y5KU).
As apparent in Figure \ref{fig:cepheus_hp}, Cepheus is composed of two components, with the near component around a distance of 350 pc, and the far component at a distance of 920 pc. We have known this for decades \citep[see, for instance][]{Grenier_1989}, but never before has the distance structure of this cloud been mapped at such a high spatial resolution. The far component dominates at $b < 14^\circ$, while the near component is seen at $b > 14^\circ$. 

\subsection{Average Cloud Distances and Uncertainties} \label{subsec:averages}

Using our new grid of distance estimates (\S \ref{subsec:pixels}), we use a Monte Carlo-based procedure to produce an average distance to each cloud. We first start by considering the full set of posterior samples (the ``chain'' returned from \texttt{dynesty}) for every pixel in the cloud boundaries which meet the minimum reddening threshold for the cloud listed in Table \ref{tab:avgtab} and described in \S\ref{subsec:pixels}. Samples are then drawn at random from ``noisy'' realization of each pixel's chain in order to incorporate sources of uncertainty within our posterior samples \citep{Speagle_2019}.\footnote{This is performed via the \texttt{dynesty} package's \texttt{simulate\_run} function. See the \texttt{dynesty} documentation on characterizing statistical and sampling errors at \href{https://dynesty.readthedocs.io/en/latest/errors.html\#combined-uncertainties}{https://dynesty.readthedocs.io/en/latest/errors.html\#combined-uncertainties}.} The random samples from each pixel are then used to compute a reddening-weighted average distance via:

\begin{equation} \label{avg_dist_eqn}
\langle d \rangle=\frac{\sum\limits_k^{n_{pix}} \langle E_{\it{Planck}} \rangle_k \; d_k}{\sum\limits_k^{n_{pix}} \langle E_{\it{Planck}} \rangle_k}
\end{equation}

\noindent where $d_k$ is the randomly-sampled ``noisy'' distance to an individual pixel and $\langle E_{\it{Planck}} \rangle_k$ is the average Planck-based reddening for the pixel. 

To characterize the statistical uncertainty in our measurements, we calculate $n=500$ realizations of $\langle d \rangle$ for each cloud. The mean and standard deviation of these five hundred average distances are then reported as the average cloud distance and its statistical uncertainty in Table \ref{tab:avgtab}. 

As discussed in \S \ref{subsec:errors}, this method does not take into account possible systematic uncertainties in the model stemming from the simplicity of our single-cloud dust model. In order to estimate the magnitude of these systematic uncertainties, we start by assuming that neighboring pixels are probing dust at the same distance. Then, for every pixel targeted in \S\ref{subsec:pixels}, we find each of its neighbors. For every target-neighbor pair, we draw ten random samples from both the target chain and the neighbor chain, and calculate the fractional differences between them relative to the target pixel. We then repeat this for every pixel in each cloud. 

\begin{figure}[t!]
\begin{center}
\includegraphics[width=0.5\columnwidth]{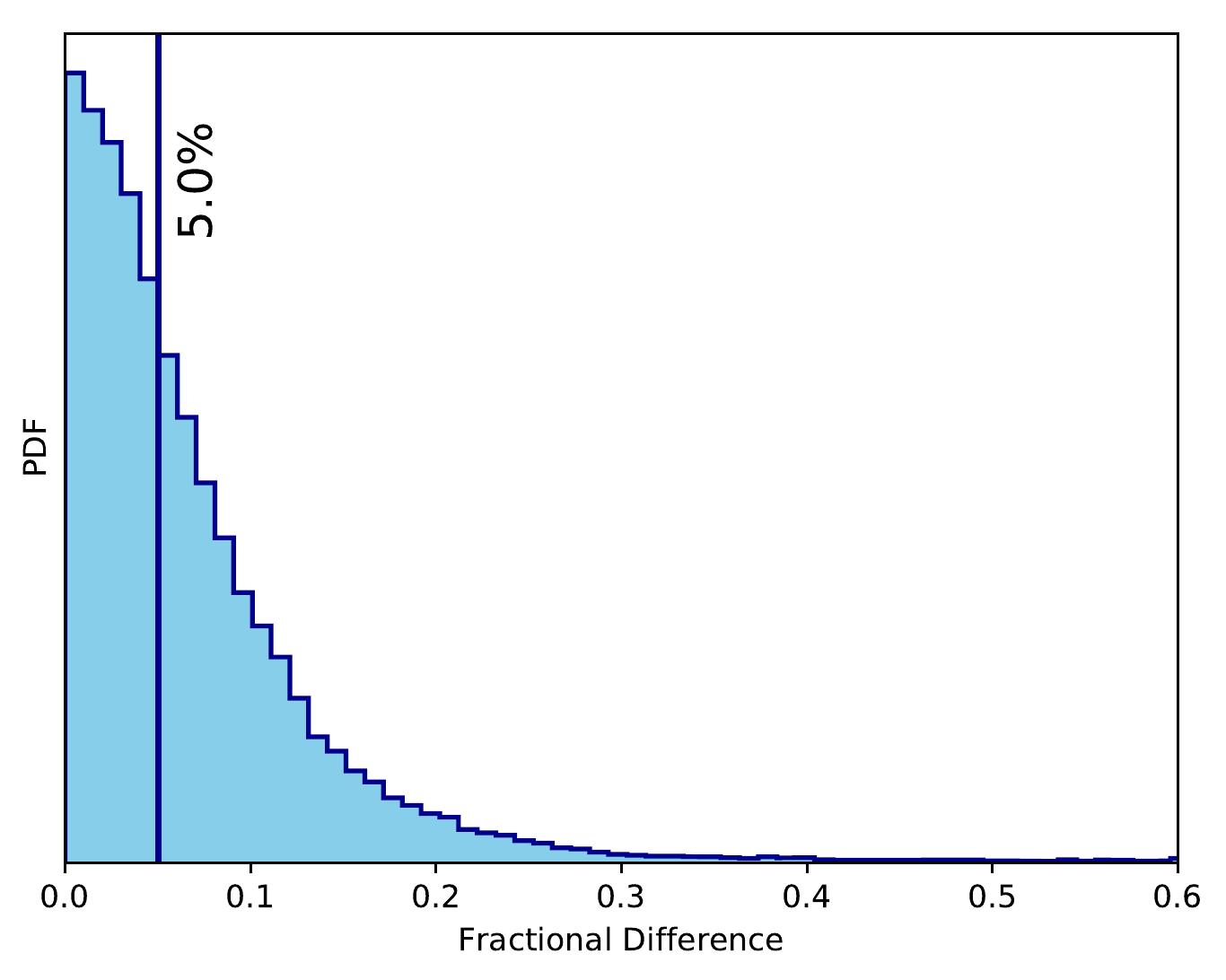}
\caption{ \label{fig:sysfig} Distribution of the fractional difference in distance between the neighboring pixels used to compute our average cloud distances in Figures \ref{fig:cepheus_hp} and \ref{fig:summary}. We find a median fractional difference of $\approx 5\%$ (blue vertical line), which we adopt as our systematic uncertainty on all reported distance measurements. See \S\ref{subsec:averages} for additional details.}
\end{center}
\end{figure}

The results from these target-neighbor pixel comparisons are shown in Figure \ref{fig:sysfig}. The distribution peaks near zero, with has a median difference of $\approx 5\%$. Based on this observed variation in neighboring pixels, we can conservatively conclude that our distance estimates are subject to a 5\% systematic uncertainty. These should be added in quadrature with our statistical uncertainty, and generally dominate the error budget. We adopt the same systematic error for all distance estimates reported in this work, but caution that the systematics could be higher for clouds at farther distances (see \S \ref{subsec:limits}).

\section{Discussion} \label{sec:discussion}

\begin{figure}[h!]
\begin{center}
\includegraphics[width=0.75\columnwidth]{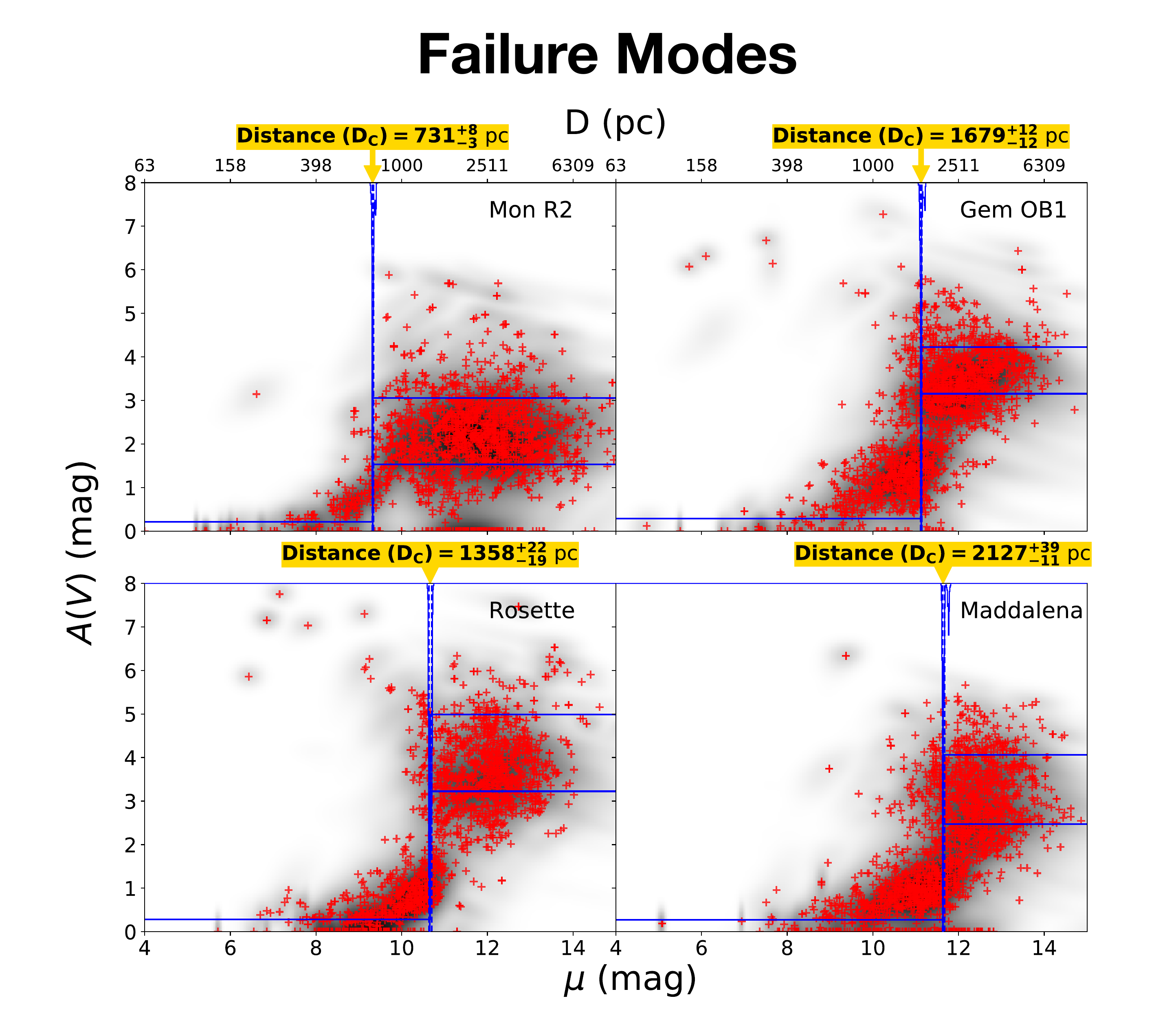}
\caption{ \label{fig:model_limitations} Same as in Figure \ref{fig:los}, but for sightlines towards Monoceros R2 ($l=215.2^\circ, b=-12.9^\circ$), Gemini OB1 ($l=193.4^\circ, b=0.3^\circ$), Rosette ($l=206.0^\circ, b=-2.7^\circ$), and Maddalena ($l=216.9^\circ, b=-2.4^\circ$). These panels highlight the limitations of our simple line-of-sight dust model. In comparison to the rest of our clouds (whose systematic uncertainty we estimate to be 5\%), these four clouds display a ``ramp" in foreground extinction leading up to the cloud. This behavior is not captured by our simple model, which assumes a flat, well-behaved foreground cloud. These modeling mismatches cause us to systematically \textit{underestimate} the distances to these sightlines and suggests higher systematic uncertainties (see \S\ref{subsec:limits}).}
\end{center}
\end{figure}

\subsection{Comparison with Distance Estimates from the Literature} \label{subsec:lit}

We have utilized a combination of optical-NIR photometry and Gaia astrometry to produce a uniform catalog of accurate distances to local molecular clouds. While other papers have used Gaia data to determine distances to molecular clouds, we argue that ours is the most comprehensive, covering a large number of clouds and taking into account larger angular areas than previous studies.

Previous approaches to estimate distances to local molecular clouds have mostly relied on constraining the distances to young stellar objects (YSOs) associated with a particular cloud, either using Gaia parallax measurements \citep{Grossschedl_2018, Dzib_2018, Ortiz_Leon_2018a, Kounkel_2018} or using VLBA trigonometric parallax measurements \citep{Ortiz_Leon_2017a, Ortiz_Leon_2017b, Ortiz_Leon_2018a, Galli_2018}. As YSOs are not included in the stellar models used to determine the distance and extinction to individual stars (see \S \ref{subsec:perstar}), they are generally poorly-fit and largely excluded from our sample following the cuts outlined in \S\ref{subsec:selection}. This means that the distance estimates obtained using YSOs provide an independent comparison to the distance estimates we present here.

Overall, we find good agreement between our distances and these YSO-derived distances. \citet{Ortiz_Leon_2017a} obtains VLBA parallax measurements towards two dark clouds in the Ophiuchus complex, L1688 and L1689, finding  distances of $137.3 \pm 1.2$ pc and $147 \pm 3.4$ pc, respectively. In our measurements of Ophiuchus, including L1688 and L1689, and the rest of the streamer extending down to $b=14^\circ$, we find an average distance of $144 \pm 7$ pc, in good agreement with their results. 

We find a similar agreement for previous measurements of the Taurus molecular cloud. \citet{Galli_2018} obtains VLBA parallax measurements towards L1495 and the B216 clump, finding distances of $129.5 \pm 0.3$ pc and $158.1 \pm 1.2$ pc, respectively. Our \href{https://faun.rc.fas.harvard.edu/czucker/Paper_Figures/Taurus.html}{distance map} for Taurus exhibit a similar bimodality, with the sightlines towards L1495 and B216 at distances of roughly $133$ pc and $152$ pc, respectively. Considering the sightlines from all pixels in Taurus, we find an average distance of $141 \pm 7$ pc, which is consistent with portions of the cloud being at the nearer distance of $\approx 130$ pc and parts of the cloud being at the farther distance of $\approx 150$ pc. 

In other regions, like the Serpens/AqR region, we find less agreement, but we attribute this in part to having targeted a more comprehensive area of the cloud compared to previous VLBA observations of YSOs. For instance, \citet{Ortiz_Leon_2017b} find distances and proper motions to seven stars in the Serpens Main and Serpens South/W40 region, and report a mean distance of $436 \pm 9.2$ pc. By contrast, we consider all area in the Serpens/AqR complex above our $E(B-V)$ threshold of 3 mag \citep[see e.g. Figure 1 in][]{Bontemps_2010}, which includes Serpens Main, Serpens South, and the W40 region, along with several other dark clouds associated with the Aquila Rift complex (towards $l=32^\circ$ and $b=3^\circ$). We find this leads to a higher overall average distance of $484 \pm 24$ pc. If we only consider the two sightlines towards Serpens Main, however, we find a lower distance of around $420$ pc, which agrees with \citet{Ortiz_Leon_2017b} given our quoted uncertainties. Like \citet{Ortiz_Leon_2017b}, we also find that the Serpens/AqR clouds form a single coherent complex, at a similar distance, with their dispersion along the line of sight $\lesssim 50$ pc.\footnote{We emphasize that all of our distance results are available online at the \href{https://dataverse.harvard.edu/dataverse/cloud_distances}{Dataverse}. Interested readers can use those results to determine what distances we computed for different parts of a cloud and possibly recompute an average for a subset of the pixels if they desire.}

The results discussed above were obtained using VLBA observations of YSOs in the pre-Gaia DR2 era, but we find similar agreement if we consider those works which use the Gaia DR2 parallax measurements. For instance, \citet{Grossschedl_2018} considered Gaia parallax measurements towards 700 YSOs in Orion A, and find a distance gradient across the cloud of $\approx 90$ pc, with the head of Orion A (towards the Orion Nebula Cluster) at a distance of around $390$ pc and the tail (towards L1647) at a distance of around $470$ pc. We find a similar trend in our data, with pixels towards the head of the cloud at a distance of $\approx 400$ pc and pixels at the tail of the cloud at distances $\approx$ 450 pc. We find an average distance to the Orion A cloud of $\approx 432 \pm 22$ pc. 

Finally, our distance estimates towards Perseus are consistent with \citet{Ortiz_Leon_2018a}, which utilized Gaia parallax observations towards the IC348 and NGC1333 star forming regions to derive distances of $320 \pm 26$ pc  and $293 \pm 22$ pc, respectively. We find a mean distance to Perseus of $294 \pm 15$ pc, which is also in strong agreement with the Gaia-informed results of \citet{Zucker_2018} ($294 \pm 17$ pc). This is is expected, as \citet{Zucker_2018} employed the same Bayesian inference framework that we do, but with a more complex line-of-sight dust model, utilizing CO velocity slices (instead of the Planck dust map) as dust templates. 

Overall, our results are in excellent agreement with previous estimates from the literature derived from independent measurements. This provides strong support for our distances, especially those we obtain for other clouds which \textit{do not} have other published Gaia-informed distances, or any other distance estimates from the literature outside of \citet{Schlafly_2014}. 

\subsection{Limitations of our Model} \label{subsec:limits}

The results presented here are based on a very simple line-of-sight dust model, characterized by a single, well-behaved foreground cloud and a single, infinitesimally-thin jump in extinction at the distance to the cloud of interest. This technique works remarkably well for a large fraction of the sample, and we believe the 5\% systematic uncertainty we quote in \S \ref{sec:results} is the characteristic systematic uncertainty across most of the full catalog of distances presented in \S \ref{sec:results}. 

Unfortunately, however, there are a few clouds where the dust possesses a more complex distribution along the line-of-sight, severely violating our model assumptions and systematically shifting the estimated distances to the clouds. This primarily affects the most distant clouds in our sample (Rosette, Maddalena, and Gemini OB1) as well as Monoceros R2. The foreground extinction towards these clouds manifests not as a discrete cloud at low extinction, but rather a steady ``ramp" up in extinction, increasing in extinction by a few magnitudes over several hundred parsecs leading up to the cloud of interest. 

This behavior is highlighted in Figure \ref{fig:model_limitations}, where we show four sightlines towards the Monoceros R2, Rosette, Maddalena, and Gemini OB1 clouds taken from our pixelated maps in \S \ref{subsec:pixels}. Because the foreground is poorly modeled in these sightlines, the default behavior is to assign low extinction to the foreground cloud ($f \approx 0.2 - 0.3$ mag) in combination with higher foreground smoothing ($s_{\rm fore}$). As a result, our fits accommodate the steady ramp up in extinction by aggressively smoothing all the individual stellar posteriors prior to the cloud distance, with typical values of $s_{fore}\approx 0.1$ ($\approx 1.2$ magnitudes of $A(V)$). This tends to result in cloud distances that are \textit{underestimated}, with at least part of the posteriors one would associate with the ramp ``by eye'' being designated as background stars. Given this behavior, we recommend that a conservative $10\%$ systematic uncertainty be adopted for Monoceros R2, Rosette, Maddalena, and Gemini OB1, with the values provided in Tables \ref{tab:avgtab} and \ref{tab:major_tab} as probable \textit{lower limits} on the distance.

\section{Conclusion} \label{sec:conclusion}
We determine distances to dozens of molecular clouds within 2 kpc from the sun using a combination of optical and near-infrared photometry and Gaia DR2 parallax measurements. Based on the broadband photometry and Gaia astrometry, we infer the distances and extinctions to tens of thousands of stars towards these clouds. We then fit these per-star distance-extinction measurements with a thin dust screen model to determine the distance to the jump in extinction corresponding to the cloud. Given the low fractional distance errors on the Gaia parallax measurements, our statistical uncertainties are small, on the order of $\approx 2-3\%$. By examining the distance correlation between line-of-sight fits towards neighboring pixels, we are able to probe the systematic uncertainty on our distances, and find that it is typically $5\%$, though it can be higher towards certain clouds due to the limitations of our very simple line-of-sight dust model. In addition to targeting over 100 sightlines towards the \citet{Magnani_1985} high-latitude clouds, we provide Gaia-informed averages distances to most of the major named clouds in the \citet{Dame_2001} CO survey, including Perseus, Orion A, Taurus, Ophiuchus, California, and Cepheus. For a subset of these clouds, we fit a grid of pixels independently over their area. This provides more detailed insight into cloud distance structure, and also provides a more comprehensive average distance determination in comparison to VLBA-based methods, which typically only target a small number of stars towards a few sightlines through each cloud. Our technique is versatile, and can be applied to almost any high-latitude sightline with an appreciable amount of extinction and an adequate number of stars. In the future we plan to incorporate deeper near-infrared data (e.g., UKIDSS, WISE) in combination with a more sophisticated line-of-sight dust model in order to probe deeper into dust-enshrouded regions and produce 3D volumetric maps of dust optimized for resolving the structure of local molecular clouds.

\acknowledgements

This work took part under the program Milky-Way-Gaia of the PSI2 project funded by the IDEX Paris-Saclay, ANR-11-IDEX-0003-02.

D.F. acknowledges support by NSF grant AST-1614941, ``Exploring the Galaxy: 3-Dimensional Structure and Stellar Streams."

This material is based upon work supported by the National Science Foundation Graduate Research Fellowship under Grant No. 1650114.

JSS is grateful to Rebecca Bleich for her patience and support and Daniel Eisenstein for his longstanding mentorship.

The Pan-STARRS1 Surveys (PS1) and the PS1 public science archive have been made possible through contributions by the Institute for Astronomy, the University of Hawaii, the Pan-STARRS Project Office, the Max-Planck Society and its participating institutes, the Max Planck Institute for Astronomy, Heidelberg and the Max Planck Institute for Extraterrestrial Physics, Garching, The Johns Hopkins University, Durham University, the University of Edinburgh, the Queen's University Belfast, the Harvard-Smithsonian Center for Astrophysics, the Las Cumbres Observatory Global Telescope Network Incorporated, the National Central University of Taiwan, the Space Telescope Science Institute, the National Aeronautics and Space Administration under Grant No. NNX08AR22G issued through the Planetary Science Division of the NASA Science Mission Directorate, the National Science Foundation Grant No. AST-1238877, the University of Maryland, Eotvos Lorand University (ELTE), the Los Alamos National Laboratory, and the Gordon and Betty Moore Foundation.

This publication makes use of data products from the Two Micron All Sky Survey, which is a joint project of the University of Massachusetts and the Infrared Processing and Analysis Center/California Institute of Technology, funded by the National Aeronautics and Space Administration and the National Science Foundation.

NOAO is operated by the Association of Universities for Research in Astronomy (AURA) under a cooperative agreement with the National Science Foundation. Database access and other data services are provided by the NOAO Data Lab.

This project used data obtained with the Dark Energy Camera (DECam), which was constructed by the Dark Energy Survey (DES) collaboration. Funding for the DES Projects has been provided by the U.S. Department of Energy, the U.S. National Science Foundation, the Ministry of Science and Education of Spain, the Science and Technology Facilities Council of the United Kingdom, the Higher Education Funding Council for England, the National Center for Supercomputing Applications at the University of Illinois at Urbana-Champaign, the Kavli Institute of Cosmological Physics at the University of Chicago, Center for Cosmology and Astro-Particle Physics at the Ohio State University, the Mitchell Institute for Fundamental Physics and Astronomy at Texas A\&M University, Financiadora de Estudos e Projetos, Funda{\c c}\~ao Carlos Chagas Filho de Amparo, Financiadora de Estudos e Projetos, Funda{\c c}\~ao Carlos Chagas Filho de Amparo \`a Pesquisa do Estado do Rio de Janeiro, Conselho Nacional de Desenvolvimento Cient\'ifico e Tecnol\'ogico and the Minist\'erio da Ci\^encia, Tecnologia e Inova{\c c}\~ao, the Deutsche Forschungsgemeinschaft and the Collaborating Institutions in the Dark Energy Survey. The Collaborating Institutions are Argonne National Laboratory, the University of California at Santa Cruz, the University of Cambridge, Centro de Investigaciones En\'ergeticas, Medioambientales y Tecnol\'ogicas–Madrid, the University of Chicago, University College London, the DES-Brazil Consortium, the University of Edinburgh, the Eidgen\"ossische Technische Hochschule (ETH) Z\"urich, Fermi National Accelerator Laboratory, the University of Illinois at Urbana-Champaign, the Institut de Ci\`encies de l'Espai (IEEC/CSIC), the Institut de F\'isica d'Altes Energies, Lawrence Berkeley National Laboratory, the Ludwig-Maximilians Universit\"at M\"unchen and the associated Excellence Cluster Universe, the University of Michigan, the National Optical Astronomy Observatory, the University of Nottingham, the Ohio State University, the University of Pennsylvania, the University of Portsmouth, SLAC National Accelerator Laboratory, Stanford University, the University of Sussex, and Texas A\&M University.

This work has made use of data from the European Space Agency (ESA) mission {\it Gaia} (\url{https://www.cosmos.esa.int/gaia}), processed by the {\it Gaia} Data Processing and Analysis Consortium (DPAC, \url{https://www.cosmos.esa.int/web/gaia/dpac/consortium}). Funding for the DPAC has been provided by national institutions, in particular the institutions participating in the {\it Gaia} Multilateral Agreement.

\software{Astropy \citep{Astropy_2018}, Dustmaps \citep{Dustmaps_2018}, Bokeh \citep{Bokeh_2018}, Healpy \citep{Healpy_2005}, dynesty \citep{dynesty_2018}, glue \citep{glueviz_2017}}

\bibliography{full_article}

\appendix
\section{Distances to Sightlines from Schlafly et al. (2014)}\label{sec:sightlines_2014}
In Tables \ref{tab:major_tab} and \ref{tab:mbm_tab} we provide updated distances to the \citet{Schlafly_2014} sightlines towards major clouds in the \citet{Dame_2001} CO survey and in the \citet{Magnani_1985} catalog, respectively.\footnote{We exclude one sightline from Schlafly et al. (2014) in Table \ref{tab:major_tab} towards the Lacerta Molecular Cloud ($l=98.7^\circ, b=-14.7^\circ$) due to the lack of foreground stars after the incorporation of the Gaia DR2 parallax measurements} In Table \ref{tab:major_tab}, we also include two additional sightlines towards the diffuse high latitude clouds Draco and the Spider, which were not targeted in \citet{Schlafly_2014}. These clouds have too low reddening to allow for the pixelated distance maps presented in \S \ref{subsec:pixels}, but targeting the most highly reddened sightlines through these clouds produces well-constrained distances to both regions. The line-of-sight extinction profiles (akin to Figure \ref{fig:los}) and the corner plots of the model parameters for all sightlines in Tables \ref{tab:major_tab} and \ref{tab:mbm_tab} are available on the \href{https://dataverse.harvard.edu/dataverse/cloud_distances}{Dataverse}. 
\newpage
\input{bigcloudcat_table.tex}

\input{mbmcloudcat_table.tex} 

\section{New NSC-2MASS Stellar Locus} \label{sec:decam_models}

To incorporate stellar modeling of the southern clouds, we derive joint NSC-2MASS stellar templates following the procedure outlined in Appendix A of \citet{Green_2015}, which describes the creation of their PS1-2MASS stellar templates. We start by querying all stars within a radial beam of $15^\circ$ centered on $(\alpha, \delta) = (356^\circ, -45^\circ$). This region was chosen due its low reddening, with a typical $E(B-V)\approx 0.01$ mag. We require that the stars have detections in all NSC-2MASS bands ($grizYJHK$), with photometric errors $<0.5$ mag in all passbands. We exclude galaxies from our sample by requiring that the source has a $>80\%$ probability of being a star (based on the NSC's \texttt{class\_star} field) and is not flagged as an extended source in 2MASS (based on the \texttt{ext\_key} field).

\begin{figure}[h!]
\begin{center}
\includegraphics[width=1.0\columnwidth]{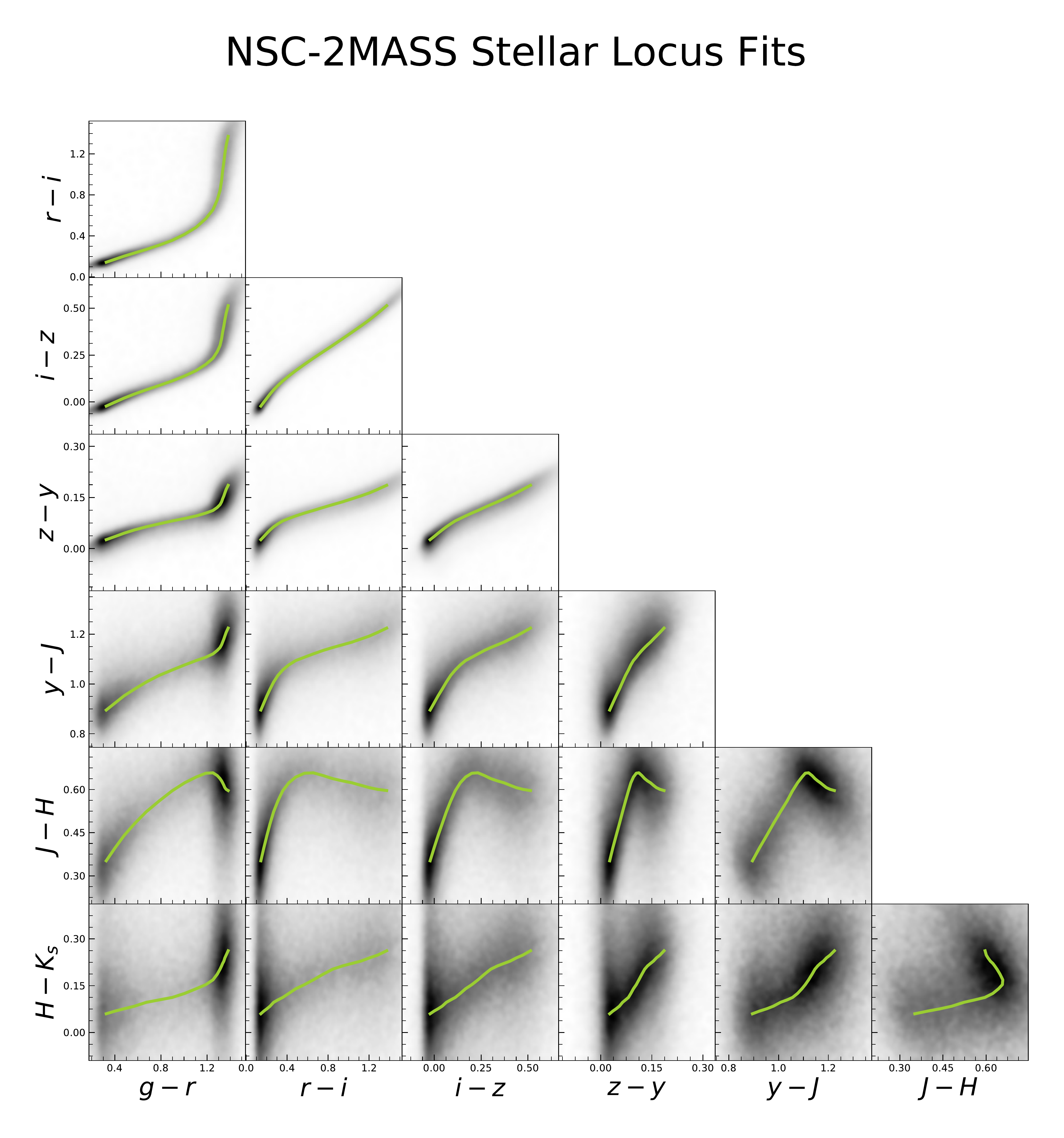}
\caption{ \label{fig:stellar_locus} The stellar locus fit used to create our new NSC-2MASS stellar templates. Each panel shows a different 2D projection of our fit in seven-dimensional color space. See Appendix \ref{sec:decam_models} for more details.}
\end{center}
\end{figure}

This results in a sample size of $\approx 200,000$ stars. After dereddening the photometry following \citet{Green_2015}, we fit a stellar locus to this sample in seven-dimensional color space following the procedure outlined in \citet{Newburg_1997}. Our resulting stellar locus is shown in Figure \ref{fig:stellar_locus}. We then associate each position along the locus with an absolute $r$ band magnitude $M_r$ and metallicity following the relations from \citet{Ivezic_2008}, which uses observations of cluster from SDSS to determine the absolute magnitudes of stars as a function of their intrinsic colors and metallicities. We finally correct for small differences in the derived $M_r^{\rm DECam}$ values from DECam and the expected $M_r^{\rm PS1}$ values from PS1 using the color corrections derived from \citet{Schlafly_2018}.

We supplement these new intrinsic NSC-2MASS stellar templates with a new reddening vector ($\mathbf{R}$) and $R(V)$ dependencies ($\mathbf{R}'$) using the procedure outlined in the Appendix of \citet{Schlafly_2016}. In brief, we integrate the dust curve provided there over the MARCS \citep{Gustafsson_2008} spectrum of a 4500 K, $[Fe/H]=0$, $\log g = 2.5$ star in the DECam $grizY$ bands. We then simultaneously set the gray component and normalization such that $A(H)/A(K)=1.55$ in the 2MASS bands and $A(V)=1$ in the Landolt $V$ band (i.e. so our reddening is in $A(V)$ units).

\section{Sampling for Our Model Parameters Using \texttt{dynesty}} \label{sec:dynesty}

We sample for our set of model parameters towards each sightline ($N$, $f$, $\mu_C$, $s_{fore}$, $s_{back}$, and $P_b$) using the nested sampling code \texttt{dynesty} (Speage 2019a, in prep). Nested sampling is similar to traditional Markov Chain Monte Carlo sampling in the sense that it can be used to generate samples from a posterior distribution. Some of the benefits of nested sampling include its ``built-in'' stopping criterion as well as its ability to estimate statistical and sampling uncertainties in a single run. 

We adopt the following \texttt{dynesty} setup:

\begin{lstlisting}
    sampler = dynesty.NestedSampler(log_likelihood, prior_transform, ndim, 
                                    bound=`multi', sample=`rwalk',
                                    update_interval=5.0, nlive=500, walks=25)
    sampler.run_nested(dlogz=0.01)
\end{lstlisting}

\noindent The \texttt{log\_likelihood} argument is the log-likelihood function described in \S \ref{subsec:los}. The \texttt{prior\_transform} argument is a function which transforms our samples from the unit cube to the target prior (a feature of the nested sampling algorithm). See \S \ref{subsec:los_priors} for our adopted priors. The \texttt{ndim} argument is the number of sampled parameters, equal to six. We set \texttt{dlogz} to 0.01, which acts as our stopping criterion. 



\end{document}

%% file: averagedist_table.tex
{
\tabletypesize{\small}
\setlength{\tabcolsep}{12pt}
\begin{deluxetable*}{ccccccc}
\tablecaption{\label{tab:avgtab} Average Distances to Local Molecular Clouds}
\tablehead{\colhead{(1)} & \colhead{(2)} & \colhead{(3)} & \colhead{(4)} & \colhead{(5)} & \colhead{(6)} & \colhead{(7)} \\ \colhead{Cloud} & \colhead{$l_{lower}$} & \colhead{$l_{upper}$} & \colhead{$b_{lower}$} & \colhead{$b_{upper}$} & \colhead{E($B-V$) Threshold} & \colhead{$D_{average}$}\\ \colhead{} & \colhead{$^\circ$} & \colhead{$^\circ$} & \colhead{$^\circ$} & \colhead{$^\circ$} & \colhead{mag} & \colhead{pc}}
\startdata
Aquila S & 35.0 & 41.0 & -20.0 & -14.0 & 0.25 & $133 \pm 2 \pm 7 $ \\
CMa OB1 & 223.0 & 227.0 & -3.0 & 0.0 & 2.0 & $1209 \pm 4 \pm 60 $ \\
California & 160.0 & 166.0 & -10.0 & -7.5 & 1.0 & $470 \pm 2 \pm 24 $ \\
Cam & 144.0 & 147.5 & 16.5 & 18.5 & 0.25 & $213 \pm 8 \pm 11 $ \\
Cepheus Far & 97.5 & 115.0 & 9.5 & 21.0 & 0.75 & $923 \pm 1 \pm 46 $ \\
Cepheus Near & 97.5 & 115.0 & 9.5 & 21.0 & 0.75 & $352 \pm 1 \pm 18 $ \\
Chamaeleon$\rm^{a}$ & 293.0 & 305.0 & -18.0 & -12.0 & 0.5 & $183 \pm 3 \pm 9 $ \\
Corona Australis & -2.0 & 3.0 & -23.0 & -16.0 & 0.25 & $151 \pm 3 \pm 8 $ \\
Crossbones & 218.5 & 220.0 & -11.0 & -7.5 & 1.0 & $886 \pm 4 \pm 44 $ \\
Gem OB1* & 188.5 & 193.5 & -1.0 & 2.5 & 2.0 & $1786 \pm 4 \pm 89 $ \\
Hercules & 42.0 & 47.0 & 7.5 & 10.0 & 0.5 & $227 \pm 1 \pm 11 $ \\
Lacerta & 94.0 & 97.0 & -12.0 & -8.0 & 0.5 & $503 \pm 5 \pm 25 $ \\
Lupus$\rm^{b}$ & 335.0 & 348.0 & 5.0 & 19.0 & 1.0 & $189 \pm 9 \pm 9 $ \\
Maddalena* & 216.0 & 219.0 & -3.0 & 1.0 & 2.0 & $2072 \pm 6 \pm 104 $ \\
Mon OB1 & 199.0 & 204.0 & 0.0 & 3.0 & 2.0 & $745 \pm 3 \pm 37 $ \\
Mon R2* & 212.0 & 216.0 & -14.0 & -11.0 & 1.0 & $778 \pm 3 \pm 39 $ \\
Ophiuchus & 351.0 & 357.0 & 13.0 & 19.0 & 1.0 & $144 \pm 2 \pm 7 $ \\
Orion A & 207.0 & 215.0 & -20.5 & -18.5 & 1.0 & $432 \pm 2 \pm 22 $ \\
Orion B & 204.0 & 209.0 & -17.0 & -13.0 & 1.0 & $423 \pm 2 \pm 21 $ \\
Orion Lam & 191.5 & 200.0 & -17.0 & -7.0 & 0.5 & $402 \pm 1 \pm 20 $ \\
Pegasus & 85.0 & 96.0 & -42.0 & -30.0 & 0.25 & $247 \pm 4 \pm 12 $ \\
Perseus & 157.5 & 161.5 & -22.0 & -16.0 & 1.0 & $294 \pm 2 \pm 15 $ \\
Polaris & 120.0 & 127.0 & 24.0 & 32.0 & 0.25 & $352 \pm 1 \pm 18 $ \\
Rosette* & 205.0 & 209.0 & -3.0 & -1.0 & 2.0 & $1304 \pm 5 \pm 65 $ \\
Serpens/AqR & 28.0 & 32.5 & 2.0 & 6.0 & 3.0 & $484 \pm 4 \pm 24 $ \\
Taurus & 166.0 & 176.0 & -19.0 & -13.0 & 1.0 & $141 \pm 2 \pm 7 $ \\
Ursa Major & 140.0 & 162.0 & 32.0 & 44.0 & 0.15 & $371 \pm 3 \pm 19 $ \\
\enddata
\tablecomments{Average distances to 27 named clouds based on Gaia astrometry and optical-NIR photometry. In (1) we list the name of the cloud. In (2)-(5) we list the longitude and latitudes bounds we consider for the clouds.  In (6) we list the $E(B-V)$ threshold used to include any pixels within this region in our distance estimate (see \S\ref{subsec:pixels}). In (7) we list the average distance we compute for the cloud given these set of pixels. The first error term is the statistical uncertainty while the second error term is the systematic uncertainty we estimate following \S\ref{subsec:averages} (5\% in distance). We recommend the uncertainties be added in quadrature. See Figure \ref{fig:summary} for an interactive distance map for each cloud, which includes all the pixels used to calculate these averages. Clouds marked with an asterisk have more complex line-of-sight dust distributions, potentially causing their distances to be underestimated (see \S \ref{subsec:limits} for more details). A machine readable version of this table is available on the \href{https://dataverse.harvard.edu/dataverse/cloud_distances}{Dataverse} (doi:10.7910/DVN/JGMNHI).}
\tablenotetext{a}{We are able to provide complete coverage of Chamaeleon I and II, but no coverage of Chamaeleon III due to a lack of multi-band optical photometry.}
\tablenotetext{b}{While we targeted pixels over the entire longitude range $335^\circ < l < 348 ^\circ$, only pixels with $l > 338^\circ$ had enough stars with multi-band optical photometry to reliably fit a distance. Thus, our average distance for Lupus is drawn from pixels near Lupus 1, 3, 5, 6, and 9, but excluding 2, 4, 7, and 8.}
\end{deluxetable*}}

%% file: bigcloudcat_table.tex
{
\newgeometry{left=0.55in,right=0.15in,bottom=0.25in,top=1.25in}
\startlongtable
\setlength{\tabcolsep}{1pt}
\tabletypesize{\footnotesize}
\begin{deluxetable*}{cccccc|cccccc}
\tablecaption{\label{tab:major_tab} Distances to Sightlines through Major Local Molecular Clouds}
\tablehead{\colhead{(1)} & \colhead{(2)} & \colhead{(3)} & \colhead{(4)} & \colhead{(5)} & \colhead{(6)} & \colhead{(1)} & \colhead{(2)} & \colhead{(3)} & \colhead{(4)} & \colhead{(5)} & \colhead{(6)}\\ \colhead{Cloud} & \colhead{$l$} & \colhead{$b$} & \colhead{$D$} & \colhead{$D_{S14}$} & \colhead{$D_{lit}$} & \colhead{Cloud} & \colhead{$l$} & \colhead{$b$} & \colhead{$D$} & \colhead{$D_{S14}$} & \colhead{$D_{lit}$} \\[-.1in] \colhead{} & \colhead{$^\circ$} & \colhead{$^\circ$} & \colhead{pc} & \colhead{pc} & \colhead{pc} & \colhead{} & \colhead{$^\circ$} & \colhead{$^\circ$} & \colhead{pc} & \colhead{pc} & \colhead{pc}}
\startdata
Aquila S & 37.8 & -17.5 & $135^{+1}_{-1} \pm 6$ & $76^{+16}_{-21} \pm 7$ &    $^{}$ & 	Ophiuchus & 352.7 & 15.4 & $139^{+2}_{-2} \pm 6$ & $116^{+7}_{-7} \pm 11$ & 119$^{10}$, 140$^{11}$ \\
Aquila S & 38.9 & -19.1 & $123^{+2}_{-3} \pm 6$ & $89^{+16}_{-22} \pm 8$ &    $^{}$ & 	Ophiuchus & 357.1 & 15.7 & $118^{+5}_{-3} \pm 5$ & $123^{+9}_{-9} \pm 12$ & 119$^{10}$, 140$^{11}$ \\
Aquila S & 36.8 & -15.1 & $143^{+3}_{-3} \pm 7$ & $125^{+7}_{-8} \pm 12$ &    $^{}$ & 	   Orion & 208.4 & -19.6 & $399^{+14}_{-7} \pm 19$ & $418^{+43}_{-34} \pm 41$ & 414$^{12}$, 400-470$^{13}$ \\
Aquila S & 39.3 & -16.8 & $128^{+3}_{-2} \pm 6$ & $111^{+12}_{-9} \pm 11$ &    $^{}$ & 	   Orion & 202.0 & -13.3 & $481^{+10}_{-13} \pm 24$ & $519^{+35}_{-34} \pm 51$ & 414$^{12}$, 400-470$^{13}$ \\
 CMa OB1 & 224.5 & -0.2 & $1262^{+6}_{-13} \pm 63$ & $1369^{+64}_{-56} \pm 136$ & 1150$^{1}$ & 	   Orion & 212.4 & -17.3 & $522^{+19}_{-54} \pm 26$ & $629^{+43}_{-43} \pm 62$ & 414$^{12}$, 400-470$^{13}$ \\
 CMa OB1 & 222.9 & -1.9 & $1169^{+21}_{-5} \pm 58$ & $1561^{+79}_{-77} \pm 156$ & 1150$^{1}$ & 	   Orion & 201.3 & -13.8 & $420^{+6}_{-9} \pm 21$ & $470^{+49}_{-33} \pm 47$ & 414$^{12}$, 400-470$^{13}$ \\
 CMa OB1 & 225.0 & -0.2 & $1266^{+3}_{-2} \pm 63$ & $1398^{+63}_{-59} \pm 139$ & 1150$^{1}$ & 	   Orion & 209.8 & -19.5 & $438^{+15}_{-26} \pm 21$ & $580^{+161}_{-107} \pm 58$ & 414$^{12}$, 400-470$^{13}$ \\
 CMa OB1 & 225.4 & 0.3 & $1268^{+1}_{-3} \pm 63$ & $1494^{+72}_{-66} \pm 149$ & 1150$^{1}$ & 	   Orion & 214.7 & -19.0 & $416^{+4}_{-5} \pm 20$ & $497^{+42}_{-36} \pm 49$ & 414$^{12}$, 400-470$^{13}$ \\
California & 163.8 & -7.9 & $466^{+17}_{-9} \pm 23$ & $431^{+33}_{-31} \pm 43$ & 450$^{2}$ & 	   Orion & 207.9 & -16.8 & $411^{+9}_{-13} \pm 20$ & $484^{+37}_{-35} \pm 48$ & 414$^{12}$, 400-470$^{13}$ \\
California & 161.2 & -9.0 & $454^{+17}_{-17} \pm 22$ & $421^{+43}_{-34} \pm 42$ & 450$^{2}$ & 	   Orion & 212.4 & -19.9 & $415^{+9}_{-16} \pm 20$ & $517^{+44}_{-38} \pm 51$ & 414$^{12}$, 400-470$^{13}$ \\
California & 162.5 & -9.5 & $436^{+10}_{-6} \pm 21$ & $377^{+39}_{-44} \pm 37$ & 450$^{2}$ & 	   Orion & 212.2 & -18.6 & $473^{+6}_{-6} \pm 23$ & $490^{+27}_{-27} \pm 49$ & 414$^{12}$, 400-470$^{13}$ \\
     Cam & 146.1 & 17.7 & $235^{+8}_{-8} \pm 11$ & $134^{+50}_{-36} \pm 13$ &    $^{}$ & 	   Orion & 209.1 & -19.9 & $445^{+25}_{-19} \pm 22$ & $478^{+84}_{-59} \pm 47$ & 414$^{12}$, 400-470$^{13}$ \\
     Cam & 144.8 & 17.8 & $220^{+11}_{-4} \pm 11$ & $208^{+37}_{-32} \pm 20$ &    $^{}$ & 	   Orion & 209.0 & -20.1 & $394^{+10}_{-9} \pm 19$ & $416^{+42}_{-36} \pm 41$ & 414$^{12}$, 400-470$^{13}$ \\
     Cam & 148.4 & 17.7 & $365^{+17}_{-9} \pm 18$ & $410^{+56}_{-86} \pm 41$ &    $^{}$ & 	   Orion & 202.0 & -14.0 & $399^{+4}_{-1} \pm 19$ & $585^{+32}_{-36} \pm 58$ & 414$^{12}$, 400-470$^{13}$ \\
     Cam & 148.8 & 17.8 & $368^{+10}_{-12} \pm 18$ & $336^{+22}_{-25} \pm 33$ &    $^{}$ & 	   Orion & 204.7 & -19.2 & $418^{+14}_{-18} \pm 20$ & $356^{+37}_{-27} \pm 35$ & 414$^{12}$, 400-470$^{13}$ \\
     Cam & 146.6 & 17.2 & $215^{+11}_{-10} \pm 10$ & $218^{+31}_{-32} \pm 21$ &    $^{}$ & 	Orion Lam & 196.7 & -16.1 & $426^{+9}_{-6} \pm 21$ & $461^{+37}_{-35} \pm 46$ & 400$^{14}$, 400$^{15}$ \\
 Cepheus & 106.4 & 17.7 & $377^{+4}_{-4} \pm 18$ & $678^{+39}_{-41} \pm 67$ & 286$^{3}$, 800-900$^{4}$ & 	Orion Lam & 194.7 & -10.1 & $425^{+24}_{-5} \pm 21$ & $378^{+31}_{-28} \pm 37$ & 400$^{14}$, 400$^{15}$ \\
 Cepheus & 110.7 & 12.6 & $989^{+4}_{-22} \pm 49$ & $859^{+33}_{-30} \pm 85$ & 286$^{3}$, 800-900$^{4}$ & 	Orion Lam & 195.5 & -13.7 & $399^{+14}_{-11} \pm 19$ & $425^{+27}_{-25} \pm 42$ & 400$^{14}$, 400$^{15}$ \\
 Cepheus & 108.3 & 12.4 & $915^{+3}_{-3} \pm 45$ & $971^{+37}_{-39} \pm 97$ & 286$^{3}$, 800-900$^{4}$ & 	Orion Lam & 194.8 & -12.1 & $423^{+11}_{-6} \pm 21$ & $463^{+38}_{-42} \pm 46$ & 400$^{14}$, 400$^{15}$ \\
 Cepheus & 105.9 & 13.8 & $951^{+6}_{-11} \pm 47$ & $961^{+29}_{-34} \pm 96$ & 286$^{3}$, 800-900$^{4}$ & 	Orion Lam & 196.9 & -8.2 & $394^{+13}_{-8} \pm 19$ & $397^{+26}_{-23} \pm 39$ & 400$^{14}$, 400$^{15}$ \\
 Cepheus & 107.0 & 9.4 & $986^{+9}_{-8} \pm 49$ & $1124^{+43}_{-43} \pm 112$ & 286$^{3}$, 800-900$^{4}$ & 	Orion Lam & 199.6 & -11.9 & $393^{+8}_{-4} \pm 19$ & $468^{+41}_{-38} \pm 46$ & 400$^{14}$, 400$^{15}$ \\
 Cepheus & 107.0 & 6.0 & $901^{+7}_{-5} \pm 45$ & $897^{+50}_{-46} \pm 89$ & 286$^{3}$, 800-900$^{4}$ & 	Orion Lam & 192.3 & -8.9 & $406^{+15}_{-17} \pm 20$ & $400^{+30}_{-29} \pm 40$ & 400$^{14}$, 400$^{15}$ \\
 Cepheus & 103.7 & 11.4 & $867^{+4}_{-8} \pm 43$ & $923^{+33}_{-35} \pm 92$ & 286$^{3}$, 800-900$^{4}$ & 	 Pegasus & 104.2 & -31.7 & $292^{+15}_{-18} \pm 14$ & $250^{+17}_{-17} \pm 25$ &    $^{}$ \\
 Cepheus & 108.4 & 18.6 & $332^{+31}_{-16} \pm 16$ & $338^{+42}_{-30} \pm 33$ & 286$^{3}$, 800-900$^{4}$ & 	 Pegasus & 105.6 & -30.6 & $256^{+15}_{-15} \pm 12$ & $210^{+25}_{-33} \pm 21$ &    $^{}$ \\
 Cepheus & 109.6 & 6.8 & $881^{+14}_{-16} \pm 44$ & $865^{+39}_{-35} \pm 86$ & 286$^{3}$, 800-900$^{4}$ & 	 Pegasus & 92.2 & -34.7 & $258^{+31}_{-99} \pm 12$ & $207^{+34}_{-36} \pm 20$ &    $^{}$ \\
 Cepheus & 108.2 & 5.5 & $891^{+10}_{-7} \pm 44$ & $853^{+48}_{-44} \pm 85$ & 286$^{3}$, 800-900$^{4}$ & 	 Pegasus & 95.3 & -35.7 & $257^{+18}_{-14} \pm 12$ & $228^{+22}_{-26} \pm 22$ &    $^{}$ \\
 Cepheus & 107.7 & 5.9 & $850^{+15}_{-26} \pm 42$ & $874^{+58}_{-52} \pm 87$ & 286$^{3}$, 800-900$^{4}$ & 	 Pegasus & 88.8 & -41.3 & $238^{+10}_{-8} \pm 11$ & $178^{+31}_{-56} \pm 17$ &    $^{}$ \\
 Cepheus & 104.0 & 9.4 & $1045^{+24}_{-9} \pm 52$ & $1045^{+40}_{-38} \pm 104$ & 286$^{3}$, 800-900$^{4}$ & 	 Perseus & 160.4 & -17.2 & $284^{+14}_{-15} \pm 14$ & $278^{+34}_{-25} \pm 27$ & 235$^{16}$, 294$^{17}$ \\
 Cepheus & 109.6 & 16.9 & $344^{+5}_{-6} \pm 17$ & $369^{+36}_{-36} \pm 36$ & 286$^{3}$, 800-900$^{4}$ & 	 Perseus & 160.7 & -16.3 & $296^{+9}_{-7} \pm 14$ & $321^{+24}_{-24} \pm 32$ & 235$^{16}$, 294$^{17}$ \\
 Cepheus & 109.0 & 7.7 & $816^{+24}_{-14} \pm 40$ & $709^{+80}_{-73} \pm 70$ & 286$^{3}$, 800-900$^{4}$ & 	 Perseus & 159.9 & -18.1 & $305^{+14}_{-20} \pm 15$ & $380^{+50}_{-96} \pm 38$ & 235$^{16}$, 294$^{17}$ \\
 Cepheus & 114.6 & 16.5 & $346^{+3}_{-4} \pm 17$ & $366^{+34}_{-32} \pm 36$ & 286$^{3}$, 800-900$^{4}$ & 	 Perseus & 158.5 & -22.1 & $243^{+12}_{-12} \pm 12$ & $266^{+27}_{-31} \pm 26$ & 235$^{16}$, 294$^{17}$ \\
 Cepheus & 113.5 & 15.9 & $336^{+3}_{-3} \pm 16$ & $389^{+22}_{-21} \pm 38$ & 286$^{3}$, 800-900$^{4}$ & 	 Perseus & 159.3 & -20.6 & $276^{+12}_{-8} \pm 13$ & $256^{+28}_{-27} \pm 25$ & 235$^{16}$, 294$^{17}$ \\
 Cepheus & 116.1 & 20.2 & $349^{+8}_{-7} \pm 17$ & $321^{+21}_{-21} \pm 32$ & 286$^{3}$, 800-900$^{4}$ & 	 Perseus & 158.6 & -19.9 & $291^{+14}_{-6} \pm 14$ & $297^{+53}_{-63} \pm 29$ & 235$^{16}$, 294$^{17}$ \\
 Cepheus & 111.8 & 20.3 & $364^{+6}_{-5} \pm 18$ & $365^{+45}_{-37} \pm 36$ & 286$^{3}$, 800-900$^{4}$ & 	 Perseus & 159.7 & -19.7 & $347^{+22}_{-24} \pm 17$ & $484^{+100}_{-121} \pm 48$ & 235$^{16}$, 294$^{17}$ \\
 Cepheus & 112.8 & 16.5 & $344^{+6}_{-9} \pm 17$ & $396^{+51}_{-53} \pm 39$ & 286$^{3}$, 800-900$^{4}$ & 	 Perseus & 159.9 & -18.9 & $279^{+18}_{-13} \pm 13$ & $297^{+43}_{-28} \pm 29$ & 235$^{16}$, 294$^{17}$ \\
 Cepheus & 108.3 & 17.6 & $346^{+11}_{-7} \pm 17$ & $372^{+44}_{-37} \pm 37$ & 286$^{3}$, 800-900$^{4}$ & 	 Perseus & 159.4 & -21.3 & $234^{+38}_{-69} \pm 11$ & $223^{+25}_{-25} \pm 22$ & 235$^{16}$, 294$^{17}$ \\
 Cepheus & 111.5 & 12.2 & $958^{+11}_{-16} \pm 47$ & $883^{+39}_{-38} \pm 88$ & 286$^{3}$, 800-900$^{4}$ & 	 Perseus & 157.8 & -22.8 & $264^{+11}_{-6} \pm 13$ & $251^{+54}_{-79} \pm 25$ & 235$^{16}$, 294$^{17}$ \\
 Cepheus & 110.1 & 17.4 & $337^{+9}_{-8} \pm 16$ & $317^{+42}_{-44} \pm 31$ & 286$^{3}$, 800-900$^{4}$ & 	 Perseus & 160.4 & -16.7 & $256^{+21}_{-44} \pm 12$ & $352^{+53}_{-50} \pm 35$ & 235$^{16}$, 294$^{17}$ \\
 Cepheus & 104.0 & 14.5 & $341^{+17}_{-13} \pm 17$ & $673^{+74}_{-86} \pm 67$ & 286$^{3}$, 800-900$^{4}$ & 	 Perseus & 160.8 & -18.7 & $285^{+17}_{-15} \pm 14$ & $232^{+53}_{-87} \pm 23$ & 235$^{16}$, 294$^{17}$ \\
 Cepheus & 103.5 & 13.5 & $359^{+4}_{-4} \pm 17$ & $372^{+36}_{-29} \pm 37$ & 286$^{3}$, 800-900$^{4}$ & 	 Perseus & 159.1 & -21.1 & $291^{+15}_{-13} \pm 14$ & $287^{+33}_{-29} \pm 28$ & 235$^{16}$, 294$^{17}$ \\
 Cepheus & 112.8 & 20.8 & $375^{+10}_{-13} \pm 18$ & $401^{+29}_{-28} \pm 40$ & 286$^{3}$, 800-900$^{4}$ & 	 Perseus & 160.8 & -17.0 & $276^{+6}_{-4} \pm 13$ & $176^{+94}_{-26} \pm 17$ & 235$^{16}$, 294$^{17}$ \\
 Cepheus & 111.5 & 20.8 & $331^{+10}_{-19} \pm 16$ & $247^{+25}_{-20} \pm 24$ & 286$^{3}$, 800-900$^{4}$ & 	 Perseus & 157.7 & -21.4 & $240^{+10}_{-11} \pm 12$ & $261^{+36}_{-43} \pm 26$ & 235$^{16}$, 294$^{17}$ \\
 Cepheus & 115.3 & 17.6 & $358^{+6}_{-5} \pm 17$ & $390^{+25}_{-24} \pm 39$ & 286$^{3}$, 800-900$^{4}$ & 	 Perseus & 158.2 & -20.9 & $287^{+7}_{-8} \pm 14$ & $288^{+39}_{-29} \pm 28$ & 235$^{16}$, 294$^{17}$ \\
 Cepheus & 107.7 & 12.4 & $961^{+6}_{-3} \pm 48$ & $957^{+34}_{-33} \pm 95$ & 286$^{3}$, 800-900$^{4}$ & 	 Perseus & 157.5 & -17.9 & $287^{+7}_{-8} \pm 14$ & $278^{+21}_{-20} \pm 27$ & 235$^{16}$, 294$^{17}$ \\
   Draco & 89.5 & 38.4 & $481^{+50}_{-45} \pm 24$ &  & 463-618$^{5}$ & 	 Perseus & 160.0 & -17.6 & $331^{+15}_{-10} \pm 16$ & $330^{+43}_{-36} \pm 33$ & 235$^{16}$, 294$^{17}$ \\
Hercules & 45.1 & 8.9 & $223^{+2}_{-2} \pm 11$ & $194^{+7}_{-7} \pm 19$ &    $^{}$ & 	 Polaris & 123.5 & 37.9 & $472^{+34}_{-37} \pm 23$ & $458^{+66}_{-75} \pm 45$ & 100$^{18}$ \\
Hercules & 44.1 & 8.6 & $223^{+3}_{-2} \pm 11$ & $184^{+5}_{-6} \pm 18$ &    $^{}$ & 	 Polaris & 129.5 & 17.3 & $341^{+19}_{-18} \pm 17$ & $337^{+166}_{-44} \pm 33$ & 100$^{18}$ \\
Hercules & 42.8 & 7.9 & $230^{+3}_{-2} \pm 11$ & $216^{+9}_{-9} \pm 21$ &    $^{}$ & 	 Polaris & 126.3 & 21.2 & $343^{+5}_{-10} \pm 17$ & $390^{+34}_{-34} \pm 39$ & 100$^{18}$ \\
 Lacerta & 96.1 & -10.2 & $504^{+6}_{-5} \pm 25$ & $517^{+27}_{-26} \pm 51$ & 520$^{6}$ & 	 Rosette & 206.8 & -1.2 & $1356^{+6}_{-15} \pm 67$ & $1540^{+69}_{-67} \pm 154$ & 1600$^{19}$ \\
 Lacerta & 95.8 & -11.5 & $473^{+5}_{-3} \pm 23$ & $509^{+29}_{-28} \pm 50$ & 520$^{6}$ & 	 Rosette & 207.8 & -2.1 & $1261^{+19}_{-12} \pm 63$ & $1383^{+85}_{-64} \pm 138$ & 1600$^{19}$ \\
Maddalena & 217.1 & 0.4 & $1888^{+22}_{-12} \pm 94$ & $2280^{+71}_{-66} \pm 228$ & 2200$^{7}$ & 	 Rosette & 205.2 & -2.6 & $1413^{+5}_{-6} \pm 70$ & $1508^{+70}_{-64} \pm 150$ & 1600$^{19}$ \\
Maddalena & 216.5 & -2.5 & $2110^{+10}_{-5} \pm 105$ & $2222^{+48}_{-47} \pm 222$ & 2200$^{7}$ & 	  Spider & 134.8 & 40.5 & $369^{+19}_{-22} \pm 18$ & &    $^{}$ \\
Maddalena & 216.8 & -2.2 & $2113^{+16}_{-8} \pm 105$ & $2071^{+59}_{-55} \pm 207$ & 2200$^{7}$ & 	  Taurus & 171.6 & -15.8 & $130^{+8}_{-7} \pm 6$ & $102^{+25}_{-32} \pm 10$ & 140$^{20}$, 130-160$^{21}$ \\
Maddalena & 216.4 & 0.1 & $2099^{+16}_{-9} \pm 104$ & $2437^{+69}_{-71} \pm 243$ & 2200$^{7}$ & 	  Taurus & 175.8 & -12.9 & $156^{+3}_{-2} \pm 7$ & $127^{+26}_{-34} \pm 12$ & 140$^{20}$, 130-160$^{21}$ \\
 Mon OB1 & 200.4 & 0.8 & $719^{+11}_{-4} \pm 35$ & $905^{+61}_{-55} \pm 90$ & 913$^{8}$ & 	  Taurus & 172.2 & -14.6 & $137^{+4}_{-2} \pm 6$ & $128^{+9}_{-10} \pm 12$ & 140$^{20}$, 130-160$^{21}$ \\
 Mon OB1 & 201.4 & 1.1 & $748^{+9}_{-10} \pm 37$ & $887^{+53}_{-44} \pm 88$ & 913$^{8}$ & 	  Taurus & 170.2 & -12.3 & $170^{+9}_{-5} \pm 8$ & $142^{+11}_{-14} \pm 14$ & 140$^{20}$, 130-160$^{21}$ \\
 Mon OB1 & 201.2 & 1.0 & $715^{+45}_{-7} \pm 35$ & $877^{+41}_{-38} \pm 87$ & 913$^{8}$ & 	  Taurus & 166.2 & -16.6 & $138^{+1}_{-1} \pm 6$ & $107^{+18}_{-20} \pm 10$ & 140$^{20}$, 130-160$^{21}$ \\
  Mon R2 & 219.2 & -7.7 & $943^{+35}_{-5} \pm 47$ & $1018^{+50}_{-43} \pm 101$ & 903$^{9}$ & 	  Taurus & 171.4 & -13.5 & $154^{+4}_{-3} \pm 7$ & $149^{+8}_{-8} \pm 14$ & 140$^{20}$, 130-160$^{21}$ \\
  Mon R2 & 215.3 & -12.9 & $767^{+12}_{-17} \pm 38$ & $788^{+34}_{-32} \pm 78$ & 903$^{9}$ & 	  Taurus & 173.5 & -14.2 & $147^{+9}_{-15} \pm 7$ & $154^{+14}_{-21} \pm 15$ & 140$^{20}$, 130-160$^{21}$ \\
  Mon R2 & 219.3 & -9.5 & $923^{+9}_{-14} \pm 46$ & $1026^{+60}_{-54} \pm 102$ & 903$^{9}$ & 	 Ursa Ma & 143.4 & 38.5 & $408^{+7}_{-4} \pm 20$ & $400^{+33}_{-27} \pm 40$ & 110$^{22}$ \\
  Mon R2 & 220.9 & -8.3 & $915^{+5}_{-3} \pm 45$ & $1052^{+35}_{-35} \pm 105$ & 903$^{9}$ & 	 Ursa Ma & 158.5 & 35.2 & $352^{+10}_{-14} \pm 17$ & $331^{+29}_{-25} \pm 33$ & 110$^{22}$ \\
  Mon R2 & 213.9 & -11.9 & $788^{+11}_{-15} \pm 39$ & $876^{+42}_{-41} \pm 87$ & 903$^{9}$ & 	 Ursa Ma & 146.9 & 40.7 & $330^{+18}_{-19} \pm 16$ & $269^{+52}_{-32} \pm 26$ & 110$^{22}$ \\
Ophiuchus & 355.2 & 16.0 & $128^{+3}_{-2} \pm 6$ & $136^{+12}_{-14} \pm 13$ & 119$^{10}$, 140$^{11}$ & 	 Ursa Ma & 153.5 & 36.7 & $352^{+19}_{-16} \pm 17$ & $353^{+38}_{-29} \pm 35$ & 110$^{22}$ \\
\enddata
\tablecomments{Distances to sightlines towards major local molecular clouds, mainly inspired by the \citet{Dame_2001} CO survey. In (1) we list the cloud associated with each sightline. In (2) and (3) we list the Galactic coordinates corresponding to the center of each sightline. In (4) we list our new Gaia-informed distances to the clouds. The first error term is the statistical uncertainty stemming from our posterior estimates (see \S \ref{subsec:los}), while the second error term is the systematic uncertainty stemming from the reliability of our line-of-sight dust model (equivalent to roughly 5\% in distance; see Figure \ref{fig:sysfig}). We recommend adding the uncertainties in quadrature. In (5) we list the distance estimates from \citet{Schlafly_2014} for comparison, including the statistical and systematic errors (roughly 10\% in distance). In (6) we compare our distances to other previous
distance estimates from the literature, which include: [1] \citet{Claria_1974}, [2] \citet{Lada_2009}, [3] \citet{Zdanavicius_2011}, [4] \citet{Grenier_1989}, [5] \citet{Gladders_1998}, [6] \citet{Kaltcheva_2009}, [7] \citet{Lee_1991}, [8] \citet{Baxter_2009}, [9] \citet{Lombardi_2011}, [10] \citet{Lombardi_2008}, [11] \citet{Ortiz_Leon_2018b}, [12] \citet{Menten_2007}, [13] \citet{Grossschedl_2018}, [14] \citet{Murdin_1977}, [15] \citet{Kounkel_2018}, [16] \citet{Hirota_2008}, [17] \citet{Zucker_2018}, [18] \citet{Zagury_1999}, [19] \citet{Blitz_1980}, [20] \citet{Kenyon_1994}, [21] \citet{Galli_2018}, [22] \citet{Penprase_1993}, respectively. See Figure \ref{fig:major_summary} for a map of these sightlines and their distances. See the \href{https://dataverse.harvard.edu/dataverse/cloud_distances}{Dataverse} (doi:10.7910/DVN/JGMNHI) for a machine readable version of this table, which also includes additional cloud parameters (see \S \ref{subsec:los}). Likewise, figures showing the cornerplots of the model parameters and the line-of-sight extinction profiles for every sightline in this table are available on the \href{https://dataverse.harvard.edu/dataverse/cloud_distances}{Dataverse} (doi:10.7910/DVN/9MZOVV).}
\restoregeometry
\end{deluxetable*}}

\newgeometry{left=0.55in,right=0.15in,bottom=0.25in,top=1.25in}
\restoregeometry

%% file: mbmcloudcat_table.tex
{\startlongtable
\setlength{\tabcolsep}{8pt}
\tabletypesize{\footnotesize}
\begin{deluxetable*}{ccccc|ccccc}
\tablecaption{\label{tab:mbm_tab} Distances to Sightlines through MBM Molecular Clouds}
\tablehead{\colhead{(1)} & \colhead{(2)} & \colhead{(3)} & \colhead{(4)} & \colhead{(5)} & \colhead{(1)} & \colhead{(2)} & \colhead{(3)} & \colhead{(4)} & \colhead{(5)} \\ \colhead{Cloud} & \colhead{$l$} & \colhead{$b$} & \colhead{$D$} & \colhead{$D_{S14}$} & \colhead{Cloud} & \colhead{$l$} & \colhead{$b$} & \colhead{$D$} & \colhead{$D_{S14}$} \\[-.1in] \colhead{} & \colhead{$^\circ$} & \colhead{$^\circ$} & \colhead{pc} & \colhead{pc} & \colhead{} & \colhead{$^\circ$} & \colhead{$^\circ$} & \colhead{pc} & \colhead{pc}}
\startdata
  1 & 110.2 & -41.2 & $265^{+42}_{-62} \pm 13$ & $228^{+45}_{-37} \pm 22$ & 	106 & 176.3 & -20.8 & $158^{+27}_{-10} \pm 7$ & $190^{+12}_{-16} \pm 19$ \\
  2 & 117.4 & -52.3 & $239^{+9}_{-7} \pm 11$ & $206^{+14}_{-12} \pm 20$ & 	107 & 177.7 & -20.4 & $141^{+4}_{-13} \pm 7$ & $197^{+12}_{-15} \pm 19$ \\
  3 & 131.3 & -45.7 & $314^{+8}_{-7} \pm 15$ & $277^{+22}_{-26} \pm 27$ & 	108 & 178.2 & -20.3 & $143^{+2}_{-2} \pm 7$ & $168^{+19}_{-21} \pm 16$ \\
  4 & 133.5 & -45.3 & $286^{+17}_{-11} \pm 14$ & $269^{+16}_{-14} \pm 26$ & 	109 & 178.9 & -20.1 & $155^{+10}_{-8} \pm 7$ & $160^{+15}_{-14} \pm 16$ \\
  5 & 146.0 & -49.1 & $279^{+16}_{-12} \pm 13$ & $187^{+86}_{-18} \pm 18$ & 	110 & 207.6 & -23.0 & $356^{+2}_{-4} \pm 17$ & $313^{+14}_{-12} \pm 31$ \\
  6 & 145.1 & -39.3 & $111^{+32}_{-31} \pm 5$ & $151^{+16}_{-66} \pm 15$ & 	111 & 208.6 & -20.2 & $400^{+4}_{-3} \pm 20$ & $366^{+9}_{-11} \pm 36$ \\
  7 & 150.4 & -38.1 & $171^{+24}_{-7} \pm 8$ & $148^{+13}_{-11} \pm 14$ & 	113 & 337.7 & 23.0 & $144^{+7}_{-7} \pm 7$ & $148^{+11}_{-13} \pm 14$ \\
  8 & 151.8 & -38.7 & $255^{+6}_{-4} \pm 12$ & $199^{+9}_{-9} \pm 19$ & 	115 & 342.3 & 24.2 & $141^{+2}_{-2} \pm 7$ & $137^{+12}_{-16} \pm 13$ \\
  9 & 156.5 & -44.7 & $262^{+22}_{-14} \pm 13$ & $246^{+42}_{-29} \pm 24$ & 	116 & 342.7 & 24.5 & $137^{+2}_{-1} \pm 6$ & $134^{+10}_{-11} \pm 13$ \\
 11 & 158.0 & -35.1 & $250^{+5}_{-7} \pm 12$ & $185^{+21}_{-20} \pm 18$ & 	117 & 343.0 & 24.1 & $138^{+2}_{-2} \pm 6$ & $140^{+9}_{-9} \pm 14$ \\
 12 & 159.4 & -34.3 & $252^{+4}_{-6} \pm 12$ & $234^{+11}_{-10} \pm 23$ & 	118 & 344.0 & 24.8 & $140^{+4}_{-4} \pm 7$ & $56^{+21}_{-17} \pm 5$ \\
 13 & 161.6 & -35.9 & $237^{+5}_{-6} \pm 11$ & $191^{+11}_{-12} \pm 19$ & 	119 & 341.6 & 21.4 & $169^{+8}_{-8} \pm 8$ & $150^{+26}_{-32} \pm 15$ \\
 14 & 162.5 & -31.9 & $275^{+3}_{-6} \pm 13$ & $233^{+11}_{-10} \pm 23$ & 	120 & 344.2 & 24.2 & $135^{+6}_{-4} \pm 6$ & $59^{+70}_{-23} \pm 5$ \\
 15 & 191.7 & -52.3 & $200^{+27}_{-24} \pm 10$ & $160^{+47}_{-64} \pm 16$ & 	121 & 344.8 & 23.9 & $140^{+5}_{-5} \pm 7$ & $118^{+11}_{-13} \pm 11$ \\
 16 & 170.6 & -37.3 & $170^{+2}_{-1} \pm 8$ & $147^{+10}_{-9} \pm 14$ & 	122 & 344.8 & 23.9 & $137^{+4}_{-3} \pm 6$ & $116^{+11}_{-19} \pm 11$ \\
 17 & 167.5 & -26.6 & $130^{+2}_{-1} \pm 6$ & $165^{+16}_{-14} \pm 16$ & 	123 & 343.3 & 22.1 & $143^{+5}_{-4} \pm 7$ & $101^{+12}_{-19} \pm 10$ \\
 18 & 189.1 & -36.0 & $155^{+3}_{-3} \pm 7$ & $166^{+18}_{-17} \pm 16$ & 	124 & 344.0 & 22.7 & $145^{+4}_{-3} \pm 7$ & $89^{+16}_{-16} \pm 8$ \\
 19 & 186.0 & -29.9 & $143^{+5}_{-2} \pm 7$ & $156^{+18}_{-18} \pm 15$ & 	125 & 355.5 & 22.5 & $129^{+5}_{-6} \pm 6$ & $115^{+16}_{-14} \pm 11$ \\
 20 & 210.9 & -36.6 & $141^{+3}_{-2} \pm 7$ & $124^{+11}_{-14} \pm 12$ & 	126 & 355.5 & 21.1 & $142^{+5}_{-6} \pm 7$ & $142^{+12}_{-17} \pm 14$ \\
 21 & 208.4 & -28.4 & $234^{+74}_{-17} \pm 11$ & $277^{+23}_{-22} \pm 27$ & 	127 & 355.4 & 20.9 & $146^{+4}_{-10} \pm 7$ & $147^{+12}_{-12} \pm 14$ \\
 22 & 208.1 & -27.5 & $266^{+30}_{-19} \pm 13$ & $238^{+27}_{-22} \pm 23$ & 	128 & 355.6 & 20.6 & $136^{+5}_{-4} \pm 6$ & $134^{+11}_{-11} \pm 13$ \\
 23 & 171.8 & 26.7 & $349^{+16}_{-14} \pm 17$ & $305^{+22}_{-22} \pm 30$ & 	129 & 356.2 & 20.8 & $139^{+4}_{-4} \pm 6$ & $141^{+11}_{-11} \pm 14$ \\
 24 & 172.3 & 27.0 & $351^{+25}_{-22} \pm 17$ & $279^{+27}_{-23} \pm 27$ & 	130 & 356.8 & 20.3 & $129^{+9}_{-3} \pm 6$ & $109^{+10}_{-13} \pm 10$ \\
 25 & 173.8 & 31.5 & $342^{+22}_{-79} \pm 17$ & $297^{+25}_{-27} \pm 29$ & 	131 & 359.2 & 21.8 & $158^{+3}_{-8} \pm 7$ & $106^{+11}_{-11} \pm 10$ \\
 27 & 141.3 & 34.5 & $376^{+9}_{-5} \pm 18$ & $359^{+23}_{-21} \pm 35$ & 	132 & 0.8 & 22.6 & $172^{+2}_{-5} \pm 8$ & $155^{+9}_{-10} \pm 15$ \\
 28 & 141.4 & 35.2 & $393^{+8}_{-9} \pm 19$ & $370^{+22}_{-20} \pm 37$ & 	133 & 359.2 & 21.4 & $161^{+3}_{-3} \pm 8$ & $98^{+12}_{-11} \pm 9$ \\
 29 & 142.3 & 36.2 & $390^{+13}_{-20} \pm 19$ & $376^{+76}_{-60} \pm 37$ & 	134 & 0.1 & 21.8 & $158^{+7}_{-3} \pm 7$ & $121^{+14}_{-28} \pm 12$ \\
 30 & 142.2 & 38.2 & $381^{+4}_{-3} \pm 19$ & $352^{+10}_{-11} \pm 35$ & 	135 & 2.7 & 22.0 & $228^{+3}_{-4} \pm 11$ & $180^{+11}_{-10} \pm 18$ \\
 31 & 146.4 & 39.6 & $395^{+7}_{-14} \pm 19$ & $325^{+27}_{-26} \pm 32$ & 	136 & 1.3 & 21.0 & $139^{+12}_{-10} \pm 6$ & $120^{+12}_{-12} \pm 12$ \\
 32 & 147.2 & 40.7 & $353^{+4}_{-8} \pm 17$ & $259^{+14}_{-15} \pm 25$ & 	137 & 4.5 & 23.0 & $169^{+13}_{-11} \pm 8$ & $146^{+11}_{-16} \pm 14$ \\
 33 & 359.0 & 36.8 & $74^{+10}_{-7} \pm 3$ & $88^{+18}_{-21} \pm 8$ & 	138 & 3.1 & 21.8 & $235^{+5}_{-7} \pm 11$ & $186^{+8}_{-8} \pm 18$ \\
 34 & 2.3 & 35.7 & $117^{+9}_{-6} \pm 5$ & $110^{+27}_{-34} \pm 11$ & 	139 & 7.6 & 24.9 & $149^{+5}_{-6} \pm 7$ & $112^{+10}_{-26} \pm 11$ \\
 35 & 6.6 & 38.1 & $86^{+17}_{-15} \pm 4$ & $89^{+17}_{-25} \pm 8$ & 	140 & 3.2 & 21.7 & $230^{+5}_{-5} \pm 11$ & $186^{+8}_{-9} \pm 18$ \\
 36 & 4.2 & 35.8 & $107^{+2}_{-2} \pm 5$ & $105^{+7}_{-7} \pm 10$ & 	141 & 4.8 & 22.6 & $163^{+5}_{-6} \pm 8$ & $127^{+13}_{-14} \pm 12$ \\
 37 & 6.1 & 36.8 & $115^{+6}_{-5} \pm 5$ & $121^{+10}_{-16} \pm 12$ & 	142 & 3.6 & 21.0 & $200^{+3}_{-4} \pm 10$ & $133^{+14}_{-13} \pm 13$ \\
 38 & 8.2 & 36.3 & $92^{+20}_{-19} \pm 4$ & $77^{+24}_{-24} \pm 7$ & 	143 & 6.0 & 20.2 & $119^{+3}_{-2} \pm 5$ & $131^{+8}_{-6} \pm 13$ \\
 39 & 11.4 & 36.3 & $102^{+4}_{-3} \pm 5$ & $94^{+15}_{-11} \pm 9$ & 	144 & 6.6 & 20.6 & $147^{+3}_{-3} \pm 7$ & $128^{+9}_{-9} \pm 12$ \\
 40 & 37.6 & 44.7 & $93^{+24}_{-20} \pm 4$ & $64^{+21}_{-25} \pm 6$ & 	145 & 8.5 & 21.8 & $108^{+6}_{-4} \pm 5$ & $152^{+19}_{-25} \pm 15$ \\
 45 & 9.8 & -28.0 & $149^{+11}_{-11} \pm 7$ & $131^{+21}_{-29} \pm 13$ & 	146 & 8.8 & 22.0 & $116^{+3}_{-9} \pm 5$ & $179^{+11}_{-12} \pm 17$ \\
 46 & 40.5 & -35.5 & $517^{+7}_{-8} \pm 25$ & $490^{+25}_{-23} \pm 49$ & 	147 & 5.9 & 20.1 & $117^{+2}_{-2} \pm 5$ & $130^{+8}_{-7} \pm 13$ \\
 47 & 41.0 & -35.9 & $491^{+6}_{-9} \pm 24$ & $475^{+25}_{-21} \pm 47$ & 	148 & 7.5 & 21.1 & $156^{+3}_{-3} \pm 7$ & $116^{+10}_{-10} \pm 11$ \\
 49 & 64.5 & -26.5 & $212^{+21}_{-35} \pm 10$ & $204^{+39}_{-33} \pm 20$ & 	149 & 7.9 & 20.3 & $164^{+3}_{-2} \pm 8$ & $114^{+13}_{-11} \pm 11$ \\
 50 & 70.0 & -31.2 & $141^{+61}_{-53} \pm 7$ & $99^{+43}_{-45} \pm 9$ & 	150 & 9.6 & 21.3 & $144^{+40}_{-26} \pm 7$ & $139^{+14}_{-12} \pm 13$ \\
 53 & 94.0 & -34.1 & $259^{+4}_{-3} \pm 12$ & $253^{+10}_{-11} \pm 25$ & 	151 & 21.5 & 20.9 & $138^{+2}_{-1} \pm 6$ & $122^{+8}_{-8} \pm 12$ \\
 54 & 91.6 & -38.1 & $245^{+4}_{-3} \pm 12$ & $231^{+11}_{-12} \pm 23$ & 	156 & 101.7 & 22.8 & $383^{+5}_{-5} \pm 19$ & $313^{+12}_{-10} \pm 31$ \\
 55 & 89.2 & -40.9 & $245^{+2}_{-3} \pm 12$ & $206^{+8}_{-6} \pm 20$ & 	157 & 103.2 & 22.7 & $358^{+3}_{-3} \pm 17$ & $325^{+11}_{-9} \pm 32$ \\
 56 & 103.1 & -26.1 & $265^{+8}_{-10} \pm 13$ & $227^{+17}_{-17} \pm 22$ & 	158 & 27.2 & -20.7 & $147^{+9}_{-8} \pm 7$ & $142^{+9}_{-10} \pm 14$ \\
 57 & 5.1 & 30.8 & $113^{+7}_{-9} \pm 5$ & $88^{+25}_{-39} \pm 8$ & 	159 & 27.4 & -21.1 & $136^{+6}_{-2} \pm 6$ & $143^{+8}_{-10} \pm 14$ \\
101 & 158.2 & -21.4 & $289^{+4}_{-6} \pm 14$ & $283^{+11}_{-10} \pm 28$ & 	161 & 114.6 & 22.5 & $360^{+2}_{-3} \pm 18$ & $308^{+8}_{-8} \pm 30$ \\
102 & 158.6 & -21.2 & $289^{+8}_{-4} \pm 14$ & $275^{+9}_{-9} \pm 27$ & 	162 & 111.7 & 20.1 & $352^{+7}_{-4} \pm 17$ & $304^{+8}_{-9} \pm 30$ \\
103 & 158.9 & -21.6 & $279^{+2}_{-1} \pm 13$ & $269^{+10}_{-9} \pm 26$ & 	163 & 115.8 & 20.2 & $336^{+3}_{-2} \pm 16$ & $293^{+6}_{-7} \pm 29$ \\
104 & 158.4 & -20.4 & $281^{+4}_{-3} \pm 14$ & $262^{+9}_{-9} \pm 26$ & 	164 & 116.2 & 20.4 & $339^{+2}_{-2} \pm 16$ & $294^{+6}_{-6} \pm 29$ \\
105 & 169.5 & -20.1 & $127^{+2}_{-2} \pm 6$ & $139^{+8}_{-9} \pm 13$ & 	165 & 116.2 & 20.3 & $341^{+1}_{-1} \pm 17$ & $291^{+7}_{-7} \pm 29$ \\
 &  &  &  & & 	            166 & 117.4 & 21.5 & $343^{+5}_{-4} \pm 17$ & $302^{+10}_{-10} \pm 30$\enddata
\tablecomments{A catalog of Gaia-informed distances to the \citet{Magnani_1985} molecular clouds. In (1) we list the MBM cloud index associated with each sightline. In (2) and (3) we list the Galactic coordinates corresponding to the center of each sightline. In (4) we list our new Gaia-informed distances to the clouds. The first error term is the statistical uncertainty stemming from our posterior estimates (see \S\ref{subsec:los}), while the second error term is the systematic uncertainty stemming from the reliability of our line-of-sight dust model (equivalent to roughly 5\% in distance; see Figure \ref{fig:sysfig}). We recommend adding the uncertainties in quadrature. In (5) we list the distance estimates from \citet{Schlafly_2014} for comparison, including the statistical and systematic errors (roughly 10\% in distance). See Figure \ref{fig:mbm_summary} for a map of these sightlines and their distances. See the \href{https://dataverse.harvard.edu/dataverse/cloud_distances}{Dataverse} for a machine readable version of this table (doi:10.7910/DVN/QW2AXB), which also includes additional cloud parameters (see \S\ref{subsec:los}). Likewise, figures showing the cornerplots of the model parameters and the line-of-sight extinction profiles for every sightline in this table are available on the \href{https://dataverse.harvard.edu/dataverse/cloud_distances}{Dataverse} (doi:10.7910/DVN/JPE8AI).}
\end{deluxetable*}}